\PassOptionsToPackage{table}{xcolor}
\documentclass[sigconf,nonacm]{acmart}

\usepackage{xcolor}
\usepackage[per-mode=symbol]{siunitx}
\DeclareSIUnit\px{px}
\newcommand{\gm}[1]{{#1}}

\newcommand{\sg}[1]{{#1}}

\DeclareRobustCommand{\ya}[1]{{#1}}
\AtBeginDocument{\pdfstringdefDisableCommands{\def\ya#1{#1}}}

\newcommand{\FpTrav}{Chunk culling}\newcommand{\fptrav}{chunk culling}
\DeclareRobustCommand{\vet}[1]{{#1}}

\usepackage{graphicx}
\usepackage{subcaption}
\usepackage{stfloats}
\usepackage{float}
\usepackage{needspace}   
\usepackage{booktabs}
\usepackage{multirow}
\usepackage{amsmath}
\usepackage{nicefrac}
\usepackage{siunitx}
\usepackage{xspace}
\usepackage[normalem]{ulem}
\usepackage{tikz}
\usetikzlibrary{arrows.meta, positioning, fit, calc}
\graphicspath{{figures/}{paper_figures/figures_out/}{paper_figures/sections/}{paper_figures/results2/}}

\AtEndPreamble{%
  \usepackage[capitalize]{cleveref}
  \crefname{section}{Sec.}{Secs.}
  \Crefname{section}{Section}{Sections}
  \crefname{table}{Tab.}{Tabs.}
  \Crefname{table}{Table}{Tables}
}
\usepackage[shortlabels,inline]{enumitem}
\setlist[itemize]{noitemsep,leftmargin=*,topsep=0em}
\setlist[enumerate]{noitemsep,leftmargin=*,topsep=0em}

\newcommand{\parhead}[1]{\paragraph{#1}}
\makeatletter
\DeclareRobustCommand\onedot{\futurelet\@let@token\@onedot}
\def\@onedot{\ifx\@let@token.\else.\null\fi\xspace}
  
\providecommand\ie{\emph{i.e}\onedot}  
\providecommand\cf{\emph{cf}\onedot}

\makeatother

\definecolor{colrtwo}{HTML}{EB6834}  
\definecolor{colck}{HTML}{1BAF7A}    
\definecolor{colckf}{HTML}{EDA100}   
\definecolor{wacvblue}{rgb}{0.21,0.49,0.74}
\DeclareRobustCommand{\rankbox}[2]{{\setlength{\fboxsep}{1pt}\colorbox{#1}{\vphantom{Ay}#2}}}

\definecolor{rankone}{HTML}{ECB4AF}    
\definecolor{ranktwo}{HTML}{EFCDA8}    
\definecolor{rankthree}{HTML}{EADE9F}  

\usetikzlibrary{backgrounds}
\let\dgsavednewcommand\newcommand
\let\newcommand\DeclareRobustCommand
\usetikzlibrary{arrows.meta, positioning, fit, calc, backgrounds}

\definecolor{dgours}{HTML}{009E73}   
\definecolor{dgink}{HTML}{222222}
\definecolor{dgmute}{HTML}{6B6B6B}
\colorlet{dgoursfill}{dgours!16!white}
\colorlet{dgadaptfill}{black!12!white}

\tikzset{
  dg/.style={font=\sffamily\scriptsize, line join=round, line cap=round,
             >={Latex[length=1.5mm, width=1.15mm]}, x=1cm, y=1cm},
  stage/.style={rectangle, rounded corners=1.8pt, align=center, inner xsep=1pt,
                inner ysep=2pt, text width=#1, minimum height=11mm, text=dgink},
  stage/.default=2.4cm,
  inherited/.style={draw=dgink, line width=0.4pt, fill=white},
  adapted/.style={draw=dgink, line width=0.6pt, fill=dgadaptfill},
  ours/.style={draw=dgours, line width=1.1pt, fill=dgoursfill},
  stepnum/.style={label={[circle, fill=dgink, text=white,
                          font=\sffamily\scriptsize\bfseries, inner sep=0pt,
                          minimum size=3.1mm, label distance=-1.7mm]north west:#1}},
  buf/.style={rectangle, draw=dgmute, line width=0.4pt, fill=white, align=center,
              inner xsep=3pt, inner ysep=1.5pt, text width=#1, minimum height=6.2mm,
              text=dgink,
              path picture={
                \draw[dgmute, line width=0.4pt]
                  ([xshift=1.3pt]path picture bounding box.north west) --
                  ([xshift=1.3pt]path picture bounding box.south west)
                  ([xshift=-1.3pt]path picture bounding box.north east) --
                  ([xshift=-1.3pt]path picture bounding box.south east);}},
  buf/.default=1.6cm,
  flow/.style={->, draw=dgink, line width=0.55pt},
  periodic/.style={flow, dash pattern=on 2.2pt off 1.6pt},
  wasted/.style={flow, line width=1.2pt},
  junction/.style={circle, fill=dgink, inner sep=0pt, minimum size=1.6pt},
  note/.style={text=dgmute, align=center, inner sep=2pt},
  panel/.style={anchor=west, inner sep=0pt, font=\sffamily\scriptsize\bfseries,
                text=dgink},
}

\newlength{\dgcardw}
\newlength{\dgcolw}
\tikzset{
  card/.style={rectangle, rounded corners=1.8pt, draw=dgink, line width=0.5pt, fill=white,
               anchor=north west, inner xsep=2pt, inner ysep=2pt, text width=\dgcardw,
               text height=1.7ex, text depth=#1, align=left, text=dgink},
  dashed card/.style={dash pattern=on 2.2pt off 1.6pt},
}

\newcommand{\pr}[2]{\makebox[\linewidth]{#1\hfill #2B}\par}                  
\colorlet{dgunit}{black!45}
\newlength{\gw}\newlength{\gcol}\newlength{\ggut}
\newsavebox{\gbx}
\newlength{\gbarh}
\definecolor{dgin}{HTML}{FBE4E1}
\definecolor{dgout}{HTML}{E0F2E4}
\tikzset{
  gw/.code={\setlength{\gw}{#1}\setlength{\gcol}{\dimexpr(#1-\ggut-5pt)/2\relax}},
  gbar/.style={path picture={\fill[#1] (path picture bounding box.north west) rectangle
                 ([yshift=-\gbarh]path picture bounding box.north east); \gfills \gdef\gfills{}}},
  gline/.style={draw=dgink, line width=0.4pt},
  gbox/.style={rectangle, rounded corners=1.8pt, draw=dgink, line width=0.5pt, fill=white,
               anchor=north west, inner xsep=2pt, inner ysep=2pt, text width=\gw,
               align=left, text=dgink, gbar=black},
  gauto/.style={rectangle, rounded corners=1.8pt, draw=dgink, line width=0.5pt, fill=white,
                anchor=north, inner xsep=3pt, inner ysep=2pt, align=left, text=dgink, gbar=black},
  gstep/.style={label={[circle, draw=dgink, line width=0.4pt, fill=white, text=dgink,
                        font=\sffamily\scriptsize\bfseries, inner sep=0pt,
                        minimum size=3.1mm, label distance=-1.7mm]north west:#1}},
}
\def\gsect{}\def\gsc{}
\newcommand{\gl}[1]{\makebox[\gw]{\textbf{\gsect}\hfill\textcolor{dgunit}{per \lowercase{#1}}}\par}
\newlength{\gy}\newlength{\gyl}\newlength{\gpadt}\newlength{\gpadb}
\def\gfills{}
\newcommand{\gfillrow}[2]{\fill[#1] ([yshift=-#2]path picture bounding box.north west) rectangle (path picture bounding box.south east);
  \ifdim#2>\gbarh \draw[gline] ([yshift=-#2]path picture bounding box.north west) -- ([yshift=-#2]path picture bounding box.north east);\fi}
\newcommand{\gfillv}[2]{\draw[gline] ([yshift=-#1]path picture bounding box.north) -- ([yshift=-#2]path picture bounding box.north);}
\newcommand{\gpanel}[1]{\ifx\gsect\empty #1\else
  \setbox\gbx=\vbox{#1}%
  \xdef\gfills{\unexpanded\expandafter{\gfills}\noexpand\gfillrow{\gsc}{\the\gy}}%
  \global\gyl=\gy
  \global\advance\gy by \dimexpr\gpadt+\gpadb+\ht\gbx+\dp\gbx\relax
  \nointerlineskip\vskip\gpadt\box\gbx\nointerlineskip\vskip\gpadb
\fi}
\newcommand{\gstrip}[1]{{\def\gsc{white}\gpanel{#1}}}
\newcommand{\gtitle}[1]{\rule[-1.9pt]{0pt}{6.1pt}\hspace{3pt}\textcolor{white}{\textbf{#1}}}
\newcommand{\gtitleu}[1]{\rule[-1.9pt]{0pt}{6.1pt}\textcolor{white}{\textbf{#1}}}
\newcommand{\gbarfix}{\vskip1.9pt}
\newcommand{\gbox}[4]{\makebox[\gw]{\gtitle{#1}\hfill\textcolor{white}{#2}}\par\gbarfix
  \global\gy=\gbarh\gdef\gfills{}%
  \gdef\gsect{Input}\gdef\gsc{dgin}#3\gdef\gsect{Output}\gdef\gsc{dgout}#4\gdef\gsect{}\vskip-0.8pt}

\newcommand{\gph}[2]{\makebox[\gw][l]{\gtitleu{#1}}\par\gbarfix\global\gy=\gbarh\gdef\gfills{}#2\gdef\gsect{}\vskip-0.8pt}
\newcommand{\gphh}[2]{\makebox[\gw]{\textbf{#1}\hfill\textcolor{dgunit}{#2}}\par}

\newcommand{\gpf}[3]{\gdef\gsect{x}\gdef\gsc{white}\gpanel{\makebox[\gw]{\textcolor{dgunit}{#1}\hfill\textcolor{dgunit}{#2}}\par #3}}
\newcommand{\gboxb}[2]{\gtitleu{#1}\\[2.4pt]\textcolor{dgmute}{#2}}
\newcommand{\pg}[3]{\gpanel{\gl{#1}\makebox[\gw][l]{\parbox[t]{\gcol}{#2}\hspace{5pt}\parbox[t]{\gcol}{#3}}\par}}
\newcommand{\pgo}[2]{\gpanel{\gl{#1}#2}}
\newcommand{\pgt}[4]{\gpanel{\makebox[\gw][l]{\parbox[t]{\gcol}{\setlength{\gw}{\gcol}\gl{#1}#2}\hspace{5pt}\parbox[t]{\gcol}{\setlength{\gw}{\gcol}\gl{#3}#4}}\par}%
  \xdef\gfills{\unexpanded\expandafter{\gfills}\noexpand\gfillv{\the\gyl}{\the\gy}}}

\newcommand{\gkeybar}[5]{{\setlength{\fboxsep}{0.8pt}\makebox[\gw][l]{\colorbox{black!22}{\makebox[\dimexpr\gw*#1/#5-1.6pt\relax]{#2}}\colorbox{black!8}{\makebox[\dimexpr\gw*#3/#5-1.6pt\relax]{#4}}}\par}}

\let\newcommand\dgsavednewcommand

\newcommand{\setcap}[3]{\expandafter\def\csname papcapT@#1\endcsname{#2}%
  \expandafter\def\csname papcapB@#1\endcsname{#3}}
\newcommand{\captitle}[1]{\csname papcapT@#1\endcsname}
\newcommand{\usecap}[1]{\captitle{#1}%
  \expandafter\ifx\csname papcapB@#1\endcsname\empty\else.\ {\normalfont\csname papcapB@#1\endcsname}\fi}
\newcommand{\inlinecap}[1]{\expandafter\def\csname papcapI@#1\endcsname{}}
\newcommand{\papfig}[2]{\begingroup
  \def\papfigtitle{#1}%
  \let\papfigcaption\caption
  \ifcsname papcapI@#2\endcsname
    \renewcommand{\caption}[2][]{\papfigcaption[##1]{##2}\Description{##1}}%
  \else\ifcsname papcapT@#2\endcsname
    \renewcommand{\caption}[2][]{\papfigcaption[\captitle{#2}]{\usecap{#2}}\Description{\captitle{#2}}}%
  \else
    \renewcommand{\caption}[2][]{\papfigcaption{\papfigtitle}\Description{\papfigtitle}}%
  \fi\fi
  \input{#2}\endgroup}

\newcommand{\brand}[1]{\textsf{#1}\xspace}
\newcommand{\method}{\brand{PennyGS}}

\inlinecap{diagrams/fig_teaser}

\inlinecap{tables/table_stage_shift}

\inlinecap{diagrams/fig_pipeline}

\inlinecap{diagrams/fig_alt_spawn_cone}

\inlinecap{diagrams/table_appearance}

\inlinecap{diagrams/fig_mech_projkeys_both}

\setcap{diagrams/fig_mech_fwd_ours}
  {Forward rasterization: each warp blends only the splats that reach it}
  {One tile of ours. Under strided pixel sampling a tile of 256 threads covers $32{\times}32$
   screen pixels and renders one in four of them each step. (1) The block loads 256 splats,
   color included, into shared memory. (2) Each splat marks which of the eight warp subtiles its
   opacity-aware bounding box reaches. (3) Each warp lists the splats it can see. (4) Each lane
   blends only that list, so splats that miss a subtile cost its warp nothing.}

\inlinecap{diagrams/fig_strided_sampling}

\inlinecap{diagrams/fig_mech_loss_pair_p1}
\inlinecap{diagrams/fig_mech_loss_pair_flow}

\inlinecap{diagrams/fig_bwd_backward}

\setcap{diagrams/table_bwd_ablation}
  {Backward rasterization}
  {Kernel time, geometric mean over \bwdAblScenes{} scenes at full resolution on an RTX~3070. Every row runs inside our pipeline on the same forward outputs, so only the backward differs; gsplat~\cite{ye2024gsplat} and Taming~\cite{mallick2024taming} are ports of the official kernels. Speedup is against gsplat.}

\inlinecap{diagrams/table_projadam_ablation_compact}

\inlinecap{tables/table_cone_cull}

\setcap{diagrams/fig_visibility_strip}
  {Most splats are not seen in a given view}
  {Share of splats that cover at least one tile, per view, over all \visViews{} views of the 15 tested scenes. Light bar: 90\% of a dataset's views; dark bar: the middle half; ticks: each scene's mean. A view sees \visMean\% of the splats on average.}

\setcap{diagrams/fig_visibility_redblue}
  {Most splats are not seen in a given view}
  {Share of splats that cover at least one tile, over every view of the 15 tested scenes. A view
   sees a third of the splats on average.}

\inlinecap{tables/table_stage_times}
\inlinecap{tables/table_sd_keys}
\inlinecap{tables/table_bwd_merged}
\setcap{tables/table_sd_bwd}
  {Backward rasterization kernels}
  {Launch configuration, achieved occupancy, share of cycles that issue an instruction, and instructions per step of
   each backward kernel on garden (Nsight Compute), with device memory on garden and stage time over \sdNumScenes{}
   scenes on an RTX~3070. gsplat and Taming-3DGS are ports of their kernels into our pipeline.}
\setcap{diagrams/fig_stage_fwd}
  {Forward rasterization, step by step}
  {Instructions per step on garden (Nsight Compute) and stage time, mean over \sdNumScenes{} scenes on an RTX~3070,
   as subtile masks, shared colour and hit lists are added. DRAM traffic is the same for all five.}
\inlinecap{diagrams/fig_quality_curve}
\setcap{tables/table_quality_ablation}
  {Quality without each change that alters training}
  {Held-out PSNR, SSIM and LPIPS (VGG) after 30k steps, mean over all \qaAllScenes{} scenes (\qaMipScenes{} Mip-NeRF~360, \qaTntScenes{} Tanks\&Temples~4K), each scene
   pooled over at least \qaRunsPerSceneMin{} runs on four GPUs. Each row puts one change back to its standard version. MB: mean splat count times float32 parameters per splat (\SI{72}{\byte} with cone fields on every splat, \SI{236}{\byte} with SH degree 3). RTX3070: training time on that GPU.}
\setcap{diagrams/fig_stage_vram}
  {Device memory of each alternative}
  {All memory the training process allocates on the GPU, on garden, split by buffer; only alternatives that change
   it are shown. Scratch: buffers a stage allocates for itself (sort storage, loss maps, Taming's checkpoints, the
   gradient buffer of separate Adam) and the ground-truth prefetch slots.}

\inlinecap{diagrams/fig_qualitative}

\inlinecap{diagrams/fig_cone_failure}

\newcommand{\cmpEntryCard}{RTX~3060}

\newcommand{\cmpOursShortEntryTime}{6m\,25s}
\newcommand{\cmpEntryRivals}{9}
\newcommand{\cmpEntryNoTime}{4}

\newcommand{\cmpGainShortMin}{3.9}
\newcommand{\cmpGainShortMax}{5.8}
\newcommand{\cmpMethods}{10}

\copyrightyear{}
\setcopyright{none} 
\acmConference[]{}{}{}
\acmBooktitle{}
\acmPrice{}
\acmDOI{} 
\acmISBN{} 
\renewcommand\footnotetextcopyrightpermission[1]{}

\acmSubmissionID{403}

\newcommand{\name}{\method}
\title[{\name}: Full Resolution 3DGS Training in Minutes on Consumer GPUs]{{\name}: Full Resolution 3DGS Training\texorpdfstring{\\}{} in Minutes on Consumer GPUs}

\author{Nicholas Jing Shou Nge$^{1}$, Yuang Shi$^{1,2,3}$, Géraldine Morin$^{2,3}$, Simone Gasparini$^{2,3}$, Anand Bhojan$^{1}$, Wei Tsang Ooi$^{1,3}$}

\affiliation{%
  \institution{$^1$National University of Singapore
  \country{Singapore}}
}
\affiliation{%
\institution{$^2$IRIT - Université de Toulouse
\country{France}}
}
\affiliation{%
\institution{$^3$IPAL, IRL2955
\country{Singapore}}
}

\begin{document}

\begin{abstract}
3D Gaussian Splatting is increasingly fed by high-resolution captures from commodity devices, yet training still happens at reduced resolution with research-grade GPUs. \gm{We \sg{present \name, a method to} achieve full-resolution (4K) training of 3DGS on consumer GPUs by improving the efficiency of every stage of a training step.} 
Improvements include a leaner splat with flat RGB color and a directional cone opacity, fused projection and key emission, forward rasterization optimizations with strided pixel sampling, an approximate tile SSIM loss, a warp-per-splat backward rasterization, and a fused backward projection and sparse Adam step. At the standard \num{30}k-step budget, \method is the fastest of \cmpMethods{} methods on Mip-NeRF~360 and Tanks\&Temples~4K on all four GPUs at competitive quality, at least \cmpGainShortMin$\times$ faster than the quickest of the others, while peaking below \SI{2}{\giga\byte} of GPU memory. 
On an \cmpEntryCard{}, a full-resolution Mip-NeRF~360 scene trains in \cmpOursShortEntryTime{}, faster than any other method trains it on an RTX~4090.
\end{abstract}



\input{diagrams/fig_teaser}

\maketitle

\section{Introduction}
\label{sec:intro}
\ifdefined\ShiftPctResRastBwd\else
\newcommand{\ShiftSplatsProj}{2.9}
\newcommand{\ShiftSplatsKeys}{1.7}
\newcommand{\ShiftSplatsSort}{2.1}
\newcommand{\ShiftSplatsRastFwd}{2.1}
\newcommand{\ShiftSplatsLoss}{1.0}
\newcommand{\ShiftSplatsRastBwd}{2.1}
\newcommand{\ShiftSplatsProjBwdAdam}{2.9}
\newcommand{\ShiftSplatsDensify}{2.9}
\newcommand{\ShiftSplatsStep}{2.3}
\newcommand{\ShiftResProj}{1.0}
\newcommand{\ShiftResKeys}{6.5}
\newcommand{\ShiftResSort}{3.8}
\newcommand{\ShiftResRastFwd}{6.3}
\newcommand{\ShiftResLoss}{12.0}
\newcommand{\ShiftResRastBwd}{4.9}
\newcommand{\ShiftResProjBwdAdam}{1.0}
\newcommand{\ShiftResDensify}{1.0}
\newcommand{\ShiftResStep}{4.3}
\newcommand{\ShiftPctSplatsProj}{7}
\newcommand{\ShiftPctSplatsKeys}{2}
\newcommand{\ShiftPctSplatsSort}{4}
\newcommand{\ShiftPctSplatsRastFwd}{7}
\newcommand{\ShiftPctSplatsLoss}{4}
\newcommand{\ShiftPctSplatsRastBwd}{38}
\newcommand{\ShiftPctSplatsProjBwdAdam}{36}
\newcommand{\ShiftPctSplatsDensify}{2}
\newcommand{\ShiftMsSplatsStep}{29.0}
\newcommand{\ShiftPctBaseProj}{5}
\newcommand{\ShiftPctBaseKeys}{3}
\newcommand{\ShiftPctBaseSort}{4}
\newcommand{\ShiftPctBaseRastFwd}{8}
\newcommand{\ShiftPctBaseLoss}{9}
\newcommand{\ShiftPctBaseRastBwd}{42}
\newcommand{\ShiftPctBaseProjBwdAdam}{28}
\newcommand{\ShiftPctBaseDensify}{2}
\newcommand{\ShiftMsBaseStep}{12.9}
\newcommand{\ShiftPctResProj}{1}
\newcommand{\ShiftPctResKeys}{4}
\newcommand{\ShiftPctResSort}{4}
\newcommand{\ShiftPctResRastFwd}{11}
\newcommand{\ShiftPctResLoss}{26}
\newcommand{\ShiftPctResRastBwd}{47}
\newcommand{\ShiftPctResProjBwdAdam}{6}
\newcommand{\ShiftPctResDensify}{0}
\newcommand{\ShiftMsResStep}{55.7}
\newcommand{\ShiftMPSplats}{1.7}
\newcommand{\ShiftSplatsPixels}{1.0}
\newcommand{\ShiftMPBase}{1.7}
\newcommand{\ShiftMPRes}{17.4}
\newcommand{\ShiftResPixels}{10.5}
\newcommand{\ShiftPctSplatsPixelStages}{56}
\newcommand{\ShiftAmdahlSplats}{1.6}
\newcommand{\ShiftPctBasePixelStages}{66}
\newcommand{\ShiftAmdahlBase}{1.7}
\newcommand{\ShiftPctResPixelStages}{92}
\newcommand{\ShiftAmdahlRes}{1.9}
\newcommand{\ShiftMsSplatsPixelStages}{16.1}
\newcommand{\ShiftMsSplatsSplatStages}{12.9}
\newcommand{\ShiftSplatsPixelStages}{1.9}
\newcommand{\ShiftAddSplatsPixelStages}{7.7}
\newcommand{\ShiftSplatsSplatStages}{2.9}
\newcommand{\ShiftAddSplatsSplatStages}{8.4}
\newcommand{\ShiftAddSplatsStep}{16.1}
\newcommand{\ShiftMsBasePixelStages}{8.5}
\newcommand{\ShiftMsBaseSplatStages}{4.4}
\newcommand{\ShiftMsResPixelStages}{51.4}
\newcommand{\ShiftMsResSplatStages}{4.3}
\newcommand{\ShiftResPixelStages}{6.1}
\newcommand{\ShiftAddResPixelStages}{42.9}
\newcommand{\ShiftResSplatStages}{1.0}
\newcommand{\ShiftAddResSplatStages}{-0.1}
\newcommand{\ShiftAddResStep}{42.8}
\fi
\ifdefined\bwdAblSpeedOurs\else
\newcommand{\bwdAblScenes}{15}
\newcommand{\bwdAblSpeedTaming}{1.27$\times$}
\newcommand{\bwdAblSpeedChunk}{2.78$\times$}
\newcommand{\bwdAblChunkOverTaming}{2.20$\times$}
\newcommand{\bwdAblStepLean}{1.14$\times$}
\newcommand{\bwdAblSpeedLean}{3.16$\times$}
\newcommand{\bwdAblSpeedOurs}{3.51$\times$}
\newcommand{\bwdAblStdStepSkip}{1.09$\times$}
\newcommand{\bwdAblStdStepWait}{1.00$\times$}
\newcommand{\bwdAblStdStepArith}{1.15$\times$}
\newcommand{\bwdAblStepSkip}{1.05$\times$}
\newcommand{\bwdAblErrFtb}{$5{\times}10^{-7}$}
\newcommand{\bwdAblStepWait}{1.05$\times$}
\newcommand{\bwdAblStepSkipWait}{1.11$\times$}
\newcommand{\bwdAblToyStepsC}{27}
\newcommand{\bwdAblToyStepsD}{25}
\newcommand{\bwdAblToyStepsE}{13}
\fi
\ifdefined\bmScenes\else
\renewcommand{\bwdAblScenes}{15}
\renewcommand{\bwdAblSpeedTaming}{1.09$\times$}
\renewcommand{\bwdAblSpeedChunk}{2.74$\times$}
\renewcommand{\bwdAblChunkOverTaming}{2.52$\times$}
\renewcommand{\bwdAblStepLean}{1.16$\times$}
\renewcommand{\bwdAblSpeedLean}{3.19$\times$}
\renewcommand{\bwdAblSpeedOurs}{3.49$\times$}
\renewcommand{\bwdAblStepSkip}{1.06$\times$}
\renewcommand{\bwdAblStepWait}{1.03$\times$}
\renewcommand{\bwdAblStepSkipWait}{1.09$\times$}
\newcommand{\bmScenes}{15}
\newcommand{\bmStepLaunch}{1.13$\times$}
\newcommand{\bmStepCkpt}{1.76$\times$}
\newcommand{\bmStepWarp}{1.27$\times$}
\newcommand{\bmTamingOcc}{33\%}
\newcommand{\bmLaunchOcc}{73\%}
\newcommand{\bmCkptInstr}{34\%}
\newcommand{\bmCkptVram}{620}
\newcommand{\bmRegsStd}{62}
\newcommand{\bmRegsLean}{48}
\newcommand{\bmOccStd}{58\%}
\newcommand{\bmOccLean}{82\%}
\newcommand{\bmLitegsSlow}{1.13$\times$}
\newcommand{\bmLitegsMin}{0.98$\times$}
\newcommand{\bmLitegsMax}{1.32$\times$}
\newcommand{\bmLitegsMaxScene}{Truck}
\newcommand{\bmLitegsOcc}{60\%}
\newcommand{\bmLitegsInstr}{1860}
\newcommand{\bmOursInstr}{2136}
\newcommand{\bmOursOcc}{83\%}
\newcommand{\bmGsplatInstr}{4261}
\newcommand{\bmInstrHalf}{2.00$\times$}
\newcommand{\bmRateOursGsplat}{1.73$\times$}
\newcommand{\bmRateTamingGsplat}{1.20$\times$}
\newcommand{\bmWarpInstr}{19\%}
\newcommand{\bmLeanInstr}{9\%}
\newcommand{\bmCullInstr}{14\%}
\fi

3D Gaussian Splatting~\cite{kerbl20233dgs} has emerged as a leading method for capturing 3D scenes for novel view synthesis and scene analysis/editing.
This opens 3D content creation to a far wider audience than research labs, from hobbyists and small studios to cultural-heritage and real-estate photographers. For these users, reconstruction has to train in their context: with their original resolution photos and computed on their consumer-level GPUs.

Current 3DGS training standards meet neither condition. Image-wise, benchmark protocols downsample training images to $1.0$--$1.6$ megapixels~\cite{barron2022mipnerf360} or shrink images wider than 1600 pixels~\cite{kerbl20233dgs}. Original captures could have up to ten times as many pixels~\cite{barron2022mipnerf360}. Hardware-wise, training speed is reported on research-grade GPUs, which have far more memory and bandwidth than consumer cards. 

Within this setting, fast trainers have cut training from hours to minutes~\cite{mallick2024taming,fastgs,chen2025dashgaussian,ye2024gsplat,dhiman2024turbogs}. Outside it, the gains largely disappear. On an \cmpEntryCard{}, \cmpEntryNoTime{} of the \cmpEntryRivals{} state-of-the-art (SOTA) trainers cannot train the full-resolution benchmark scenes in practice, and even the fastest of the others needs over half an hour per scene (\cref{tab:main_comparison}).

We profile a training step while varying the resolution and the number of splats (\cref{tab:stage-shift}), and find that two pressures build up.
The first is \emph{pixels}. With \ShiftResPixels$\times$ the pixels, the step becomes \ShiftResStep$\times$ slower.
The second is \emph{memory traffic}. The stages that work per splat do little arithmetic but move many bytes, so they grow with the number of splats.
Achieving gains robust to splat count and resolution changes therefore requires every stage to become cheaper, instead of focusing on speeding up a single stage. 

Our key insight is that redundancy still exists throughout training pipeline in three forms. Redundant neighboring pixels of a high-resolution photo are nearly alike, yet every step renders, compares, and back-propagates all of them, and SSIM makes pixels enter $121$ overlapping windows. 
Redundant bytes of splats not only incurs a larger final model size, but puts load on memory bandwidth bound stages. Standard Adam optimizer also updates them even for the two-thirds of the splats a view does not see.
The third is redundant passes. Tile overlaps are enumerated twice to size the sort buffer, and intermediate maps and gradients are written to memory only to be read back by the next kernel. \cref{sec:overview} analyzes each form in detail.

\ya{Based on this insight, we propose \method{}, a 3DGS training system that removes redundant work from every stage of a training step, so that full-resolution scenes train in minutes on consumer GPUs.}
Removing redundancy naively, however, would lose what it provides, \ie, fine detail, view-dependent appearance, and supervision of every pixel. 
Several of the removals below already exist, often only for rendering or on their own; \method{} adapts them to training and combines them with new ones, so that every stage of the step becomes cheaper while all three are kept (\cref{fig:pipeline}). 
For pixels, each step supervises a strided quarter of the actual full-resolution pixels, as in the dilated rendering of Turbo-GS~\cite{dhiman2024turbogs}, and we cycle the offset so that no pixel is averaged away and every pixel is still supervised once every four steps (\cref{sec:sampling}); likewise, SSIM is evaluated over non-overlapping tiles whose origin shifts at every step, so that no tile boundary stays fixed (\cref{sec:loss}). For bytes, a splat carries a flat RGB color, as in UBS~\cite{liu2026ubs}, and a directional cone opacity in the spirit of ARF~\cite{yang2026arf}, in $14$ or $18$ parameters instead of $59$; unlike ARF, the cone lets projection drop a splat before it emits any keys (\cref{sec:appearance}). Adam updates only the splats a view sees, following Taming 3DGS~\cite{mallick2024taming}, over compact moments as in LichtFeld~\cite{lichtfeld} (\cref{sec:adam}). For passes, projection and key emission run as one kernel with no round trip to the host, which RoofGS~\cite{luo2026roofgs} does for rendering (\cref{sec:projkeys}); the tile SSIM obtains its gradient in closed form and runs the whole loss in a single pass; backward projection is fused with the Adam update, as in VkSplat~\cite{vksplat} and LichtFeld~\cite{lichtfeld}, and densification reuses the renders that training already produces (\cref{sec:densify}). Finally, we restructure the backward rasterization. As in LiteGS~\cite{liao2026litegs}, a warp works on one splat at a time, but it walks only the parts of a tile the splat reaches and several splats are in flight per tile, which makes it \bwdAblSpeedOurs{} faster than that of gsplat~\cite{ye2024gsplat} (\cref{sec:bwd}).

In summary, our contributions are:
\begin{itemize}
    \item \emph{A stage-level characterization} of 3DGS training across resolution and splat count, which shows that full-resolution training on consumer GPUs is dominated by per-pixel work and per-splat memory traffic, much of it redundant.
    \item \emph{A training system that removes this redundancy at every stage}. Its main new components are fused projection and key emission, a single-pass tile SSIM and a per-chunk backward rasterization. It combines them with a lean splat representation and with adapted techniques for pixel sampling, densification and sparse Adam.
    \item \emph{A full-resolution evaluation} of \cmpMethods{} SOTAs trainers on four GPUs, from an \cmpEntryCard{} to an RTX~4090, over $15$ scenes at 4K. On an \cmpEntryCard{}, our system trains a full-resolution Mip-NeRF~360 scene in \cmpOursShortEntryTime{}, faster than any other method trains it on an RTX~4090. More broadly, at the standard \num{30}k-step budget it is the fastest of the SOTAs on all four GPUs and datasets, at least \cmpGainShortMin$\times$ faster than the quickest of the others, while peaking below \SI{2}{\giga\byte} of GPU memory and keeping quality competitive.
\end{itemize}

The remainder of this paper is organized as follows. \cref{sec:related} reviews prior work on accelerating 3DGS training, on training at high resolution and large scale, and on compact representations and memory. \cref{sec:method} first profiles a training step and locates its redundant work (\cref{sec:overview}), then presents our design stage by stage. \cref{sec:results} compares \method{} with \cmpEntryRivals{} SOTA trainers on four consumer GPUs, then isolates the contribution of each change and examines its trade-off between speed and quality. Finally, \cref{sec:conclusion} concludes the paper and discusses limitations and future work. 

We will release the code and scripts upon acceptance.

\section{Related Work} \label{sec:related}
\papfig{Data movement per stage, ours}{diagrams/fig_pipeline}
\parhead{Accelerating 3DGS training.}
Several approaches can speed up 3DGS training. Some works revise the densification schedule and criteria, including steering fixed primitive budgets, replacing the positional gradient heuristic, or regulating growth by error~\cite{mallick2024taming,fastgs,fang2024minisplatting2,dhiman2024turbogs,chen2025dashgaussian,zhang2024pixelgs,bulo2024revising,grubert2025improving,elrawy2025opacity,gao2025easysplat}, reformulating densification as sampling under a fixed budget~\cite{kheradmand20243dgs}, or training on primitive subsets~\cite{wang2024group}.  
Others replace or strengthen the optimizer with second-order or Levenberg--Marquardt steps~\cite{hollein2024gslm,lan2025gs2,zhang2026fastconverging}. A third family re-engineers the differentiable rasterizer itself: faster kernels and gradient accumulation~\cite{ye2024gsplat,durvasula2023distwar,feng2024flashgs,yuan2025hardware}, and revised or removed depth sorting~\cite{radl2024stopthepop,kheradmand2025stochasticsplats,hou2024sortfree,hahlbohm2024hybrid}. Speedy-Splat~\cite{hanson2025speedysplat} tightens the pixel set each splat touches. \vet{Among trainers re-engineered for speed, VkSplat~\cite{vksplat} and LichtFeld~\cite{lichtfeld} fuse backward projection with the Adam update, LichtFeld over compact moments, and LiteGS~\cite{liao2026litegs} puts a warp on one splat at a time in the backward pass. Taming-3DGS~\cite{mallick2024taming} updates only the splats a view sees. For rendering, RoofGS~\cite{luo2026roofgs} runs projection and key emission as one kernel.}
\ya{These methods share an implicit assumption: their speedups are designed and measured at benchmark resolution, where per-pixel stages are a smaller share of the step. Most of them also target one part of training: primitive count, the number of steps, or a single kernel.}

\parhead{Resolution and scale.}
Standard benchmark protocols supervise at a fraction of the captured resolution. Work that does confront resolution falls into three groups. Anti-aliasing methods make rendering consistent across scales without reducing the cost of training at high resolution~\cite{yu2024mipsplatting,yan2023multiscale}. Scale-out systems distribute the problem over multiple GPUs or spatial partitions~\cite{zhao2025grendel,kerbl2024hierarchical,lin2024vastgaussian,liu2024citygaussian,li2025hrgs}.
Finally, several accelerators exploit reduced resolution as scaffolding during optimization~\cite{chen2025dashgaussian}. Turbo-GS~\cite{dhiman2024turbogs} renders a chessboard subset of pixels, drawing one offset at random per iteration and treating the result as a downsampled image.
The third group points to an interesting idea: at high resolution neighboring pixels are largely redundant, so a training step need not see every pixel. 

\papfig{Per-stage timings}{tables/table_stage_shift}
\parhead{Compact representations and memory.}
The reference implementation reports peak training memory above \SI{20}{\giga\byte} on a \SI{48}{\giga\byte} A6000~\cite{kerbl20233dgs}, which motivated training within tighter budgets. Compact representations and pruned or proxy-structured models reduce parameters during training~\cite{girish2024eagles,lee2024compact,papantonakis2024reducing,gao2025proxygs}; gsplat~\cite{ye2024gsplat} reports ${\sim}4\times$ lower training memory, and offloading systems move Gaussians to host memory to scale training~\cite{lee2025gsscale} and rendering~\cite{zhao2026clm} beyond device VRAM.
\ya{These methods treat memory capacity as the main constraint, and often spend extra time to meet it, for example on host--device transfers. On consumer GPUs, we find that the per-splat stages are bound by the bytes they move rather than by computation (\cref{sec:overview}), suggesting memory bandwidth limits training just as much. A compact splat therefore pays off twice, in memory and in speed. \vet{Yet appearance takes most of a splat's parameters, and the models that revisit it mostly add to them: Spec-Gaussian~\cite{yang2024specgaussian} and Spherical Voronoi~\cite{disario2025sphericalvoronoi} extend or replace SH with other directional color bases, and UBS~\cite{liu2026ubs} and ARF~\cite{yang2026arf} make opacity view-dependent, at 24 to 60 appearance parameters per splat (\cref{tab:appearance}). Scaffold-GS~\cite{lu2024scaffoldgs} instead predicts color and opacity from anchor features with small MLPs, trading parameters for computation.} This motivates our lean splat representation (\cref{sec:appearance}).}

\ifdefined\bwdAblSpeedOurs\else
\newcommand{\bwdAblScenes}{15}
\newcommand{\bwdAblSpeedTaming}{1.27$\times$}
\newcommand{\bwdAblSpeedChunk}{2.78$\times$}
\newcommand{\bwdAblChunkOverTaming}{2.20$\times$}
\newcommand{\bwdAblStepLean}{1.14$\times$}
\newcommand{\bwdAblSpeedLean}{3.16$\times$}
\newcommand{\bwdAblSpeedOurs}{3.51$\times$}
\newcommand{\bwdAblStdStepSkip}{1.09$\times$}
\newcommand{\bwdAblStdStepWait}{1.00$\times$}
\newcommand{\bwdAblStdStepArith}{1.15$\times$}
\newcommand{\bwdAblStepSkip}{1.05$\times$}
\newcommand{\bwdAblErrFtb}{$5{\times}10^{-7}$}
\newcommand{\bwdAblStepWait}{1.05$\times$}
\newcommand{\bwdAblStepSkipWait}{1.11$\times$}
\newcommand{\bwdAblToyStepsC}{27}
\newcommand{\bwdAblToyStepsD}{25}
\newcommand{\bwdAblToyStepsE}{13}
\fi
\ifdefined\bmScenes\else
\renewcommand{\bwdAblScenes}{15}
\renewcommand{\bwdAblSpeedTaming}{1.09$\times$}
\renewcommand{\bwdAblSpeedChunk}{2.74$\times$}
\renewcommand{\bwdAblChunkOverTaming}{2.52$\times$}
\renewcommand{\bwdAblStepLean}{1.16$\times$}
\renewcommand{\bwdAblSpeedLean}{3.19$\times$}
\renewcommand{\bwdAblSpeedOurs}{3.49$\times$}
\renewcommand{\bwdAblStepSkip}{1.06$\times$}
\renewcommand{\bwdAblStepWait}{1.03$\times$}
\renewcommand{\bwdAblStepSkipWait}{1.09$\times$}
\newcommand{\bmScenes}{15}
\newcommand{\bmStepLaunch}{1.13$\times$}
\newcommand{\bmStepCkpt}{1.76$\times$}
\newcommand{\bmStepWarp}{1.27$\times$}
\newcommand{\bmTamingOcc}{33\%}
\newcommand{\bmLaunchOcc}{73\%}
\newcommand{\bmCkptInstr}{34\%}
\newcommand{\bmCkptVram}{620}
\newcommand{\bmRegsStd}{62}
\newcommand{\bmRegsLean}{48}
\newcommand{\bmOccStd}{58\%}
\newcommand{\bmOccLean}{82\%}
\newcommand{\bmLitegsSlow}{1.13$\times$}
\newcommand{\bmLitegsMin}{0.98$\times$}
\newcommand{\bmLitegsMax}{1.32$\times$}
\newcommand{\bmLitegsMaxScene}{Truck}
\newcommand{\bmLitegsOcc}{60\%}
\newcommand{\bmLitegsInstr}{1860}
\newcommand{\bmOursInstr}{2136}
\newcommand{\bmOursOcc}{83\%}
\newcommand{\bmGsplatInstr}{4261}
\newcommand{\bmInstrHalf}{2.00$\times$}
\newcommand{\bmRateOursGsplat}{1.73$\times$}
\newcommand{\bmRateTamingGsplat}{1.20$\times$}
\newcommand{\bmWarpInstr}{19\%}
\newcommand{\bmLeanInstr}{9\%}
\newcommand{\bmCullInstr}{14\%}
\fi
\newcommand{\paAblgsplatms}{10.87}
\newcommand{\paAblgsplatSpeed}{1.00$\times$}
\newcommand{\paAblgsplatmsRaw}{10.87}
\newcommand{\paAblVkSplatms}{5.67}
\newcommand{\paAblVkSplatSpeed}{1.92$\times$}
\newcommand{\paAblVkSplatmsRaw}{5.67}
\newcommand{\paAblLichtFeldms}{8.78}
\newcommand{\paAblLichtFeldSpeed}{1.24$\times$}
\newcommand{\paAblLichtFeldmsRaw}{8.78}
\newcommand{\paAblAms}{2.41}
\newcommand{\paAblASpeed}{4.51$\times$}
\newcommand{\paAblBms}{1.76}
\newcommand{\paAblBSpeed}{6.19$\times$}
\newcommand{\paAblCms}{0.67}
\newcommand{\paAblCSpeed}{16.25$\times$}
\newcommand{\paAblVisible}{34\%}

\section{Methodology}
\label{sec:method}

\subsection{\ya{Motivation} and Overview}
\label{sec:overview}

\label{sec:pipeline}

\ifdefined\ShiftPctResRastBwd\else\fi
\newcommand{\stScenes}{15}
\newcommand{\stOursLossMs}{0.23}
\newcommand{\stOursPsnr}{26.48}
\newcommand{\stOursSsim}{0.832}
\newcommand{\stOursLpips}{0.287}
\newcommand{\stLossStdLossMs}{1.99}
\newcommand{\stLossStdLossX}{8.5}
\newcommand{\stLossStdStepX}{1.16}
\newcommand{\stLossStdStepPct}{-13.5}
\newcommand{\stLossStdPsnr}{26.36}
\newcommand{\stLossStdSsim}{0.833}
\newcommand{\stLossStdLpips}{0.290}
\newcommand{\stLossStdPsnrGain}{0.12}
\newcommand{\stSHThreeLossMs}{0.23}
\newcommand{\stSHThreeLossX}{1.0}
\newcommand{\stSHThreeStepX}{0.98}
\newcommand{\stSHThreeStepPct}{2.5}
\newcommand{\stSHThreePsnr}{25.99}
\newcommand{\stSHThreeSsim}{0.829}
\newcommand{\stSHThreeLpips}{0.293}
\newcommand{\stSHThreePsnrGain}{0.49}
\newcommand{\stFwdStdLossMs}{0.94}
\newcommand{\stFwdStdLossX}{4.0}
\newcommand{\stFwdStdStepX}{3.35}
\newcommand{\stFwdStdStepPct}{-70.2}
\newcommand{\stFwdStdPsnr}{26.49}
\newcommand{\stFwdStdSsim}{0.832}
\newcommand{\stFwdStdLpips}{0.287}
\newcommand{\stFwdStdPsnrGain}{-0.01}
\newcommand{\stTabOursPsnr}{26.48}
\newcommand{\stTabOursSsim}{0.832}
\newcommand{\stTabFwdStdPsnr}{26.50}
\newcommand{\stTabFwdStdSsim}{0.832}
\newcommand{\stTabFwdStdDPsnr}{+0.02}
\newcommand{\stTabFwdStdDSsim}{0.000}

\newcommand{\coneCullOfVisible}{32}
\newcommand{\coneCullOfVisibleMin}{15}
\newcommand{\coneCullOfVisibleMax}{50}
\newcommand{\coneCullOfAll}{17}
\newcommand{\coneCullOfCone}{34}
\newcommand{\coneShare}{48}
\newcommand{\coneCullScenes}{15}
\newcommand{\coneInViewPct}{64}
\newcommand{\baseInViewPct}{38}
\newcommand{\coneCullOfConeInView}{52}
\newcommand{\coneCullPairs}{40}
\providecommand{\visMean}{33}
\providecommand{\visPfive}{15}
\providecommand{\visPninetyfive}{53}
\providecommand{\visSceneMin}{21}
\providecommand{\visSceneMax}{50}
\providecommand{\visViews}{3960}

\ya{
\parhead{One training step.}
A step renders one training view, compares it with the ground truth, and updates the splats (\cref{fig:pipeline}). Different stages scale differently:
\begin{itemize}
    \item \emph{Per-splat stages} visit every splat once. Forward projection maps each 3D Gaussian to a 2D splat on the image plane; backward projection maps the 2D gradients back to the 3D parameters and Adam~\cite{kingma2015adam} updates them.
    \item \emph{Per-pixel stages} scale only with pixels. Loss (L1 and SSIM) compares $C$ with the ground truth and returns $\partial L/\partial C$, and is the only stage that is purely per-pixel.
    \item Most stages however scale with both. Key emission writes a key for every tile a splat overlaps, sort orders the keys by tile and depth, forward rasterization blends each tile's sorted splats, and backward rasterization computes gradients of 2D splats from the per-pixel losses.
\end{itemize}
Periodic densification also runs alongside.

\parhead{Where the time goes.}
\cref{tab:stage-shift} profiles a training step in VkSplat~\cite{vksplat}. In a example basline configuration with images \num{1600} pixels wide and 1M splats (as in MCMC~\cite{kheradmand20243dgs}), backward rasterization takes \ShiftPctBaseRastBwd\% of the step and backward projection with Adam \ShiftPctBaseProjBwdAdam\%. Varying splats and resolution shows how stages scale unevenly:
\begin{itemize}
    \item \emph{More splats} (3M) slow the per-splat stages the most, \ShiftSplatsProj$\times$ against \ShiftSplatsStep$\times$ for the whole step, and backward projection with Adam grows to \ShiftPctSplatsProjBwdAdam\%. These stages are bound by memory bandwidth: they load and store dozens of bytes per splat, its parameters and Adam moments, with little arithmetic.
    \item \emph{Full resolution} ($5187{\times}3361$, \ShiftResPixels$\times$ the pixels) leaves the per-splat stages unchanged but inflates everything that touches pixels: key emission grows \ShiftResKeys$\times$, as each splat now overlaps more tiles, forward rasterization \ShiftResRastFwd$\times$, and the loss \ShiftResLoss$\times$, to \ShiftPctResLoss\% of the step. The loss grows fastest because SSIM reads an $11{\times}11$ window around every pixel and takes two passes over the image (\cref{sec:loss}).
\end{itemize}

\parhead{Our design.}
Full-resolution training on a consumer GPU meets all three pressures at once: millions of splats, ten times the pixels, and limited bandwidth and compute. No single stage is the bottleneck, so we rework every stage, guided by two principles: \emph{move fewer bytes per splat} in the bandwidth-bound per-splat and per-pair stages, and \emph{do less work per pixel} in the compute-bound per-pixel stages.

\cref{fig:pipeline} shows what each stage of our pipeline reads and writes, and \cref{tab:stage_times} gives its per-stage times.
}

\subsection{Directional Opacity and RGB Color}
\label{sec:appearance}

The opacity $o$ is an important parameter in 3DGS. It is the criterion for pruning, which removes splats with value $o<0.05$. Opacity also indirectly impacts splat coverage/size, due to cutoff of Gaussian influence below a certain alpha, which is opacity times the Gaussian falloff at the pixel. Implementations even go out of their way to interfere with it. 3DGS~\cite{kerbl20233dgs} resets every splat's opacity to at most $0.01$ every 3k iterations, with the Adam moments of opacity zeroed. In LichtFeld Studio~\cite{lichtfeld}, instead of a reset, opacity decays.

\papfig{Cost of appearance per splat per view}{diagrams/table_appearance}
Despite this, opacity is still only one parameter (\SI{4}{\byte}), while much more of a splat's parameter budget is allocated for directional color. Various works have offered more compact representations for directional color, and others explored allocating more parameters toward directional opacity as well (\cref{tab:appearance}).

While not conceptually novel, we find no works specifically recommending a simple flat RGB color with an equivalently lightweight directional color. The closest, ARF~\cite{yang2026arf}, uses $5$ extra parameters $(d, T, \beta)$ on top of opacity to model directional opacity while keeping SH3 color.

\papfig{Cone spawn and cone opacity}{diagrams/fig_alt_spawn_cone}

\parhead{Directional opacity.} We model directional opacity $\hat{o}$ with a normal-like cone axis $\mathbf{n}$ (3 fp32, unit length) and linear falloff $k$ (fp32) in $\cos\theta$, in addition to the regular opacity parameter $o$. \sg{Our directional opacity $\hat{o}$ is then computed as} 
\begin{equation}
  \hat{o} \;=\; o \cdot \operatorname{saturate}\!\big(1 - k\,(1 - \mathbf{w}\cdot\mathbf{n})\big),
  \label{eq:cone}
\end{equation}
\sg{where $\mathbf{w}$ is the unit direction from the splat to the camera and $\theta$ is the angle between $\mathbf{w}$ and $\mathbf{n}$ (so $\cos\theta = \mathbf{w}\cdot\mathbf{n}$, \cf \cref{fig:alt-spawn-cone}, right), and $\operatorname{saturate}$ clamps to $[0,1]$.}
With $k = 0$ the splat is visible from every direction, as in 3DGS; otherwise $\hat{o}$ falls to zero at $\cos\theta = 1 - 1/k$, so a larger $k$ gives a narrower cone (\cref{fig:alt-spawn-cone}, right). 

The above is computationally cheap. We pair this with a direct RGB color of 3 floats, one per channel, in place of SH.

\parhead{Culling.}  Because $\hat{o} = 0$ for all views outside the cone ($\cos\theta \le 1 - 1/k$), the splat can be culled in projection and excluded from future stages, similar to back-face culling in traditional graphics.

\parhead{Adding cone splats.} During training, base splats with no cones (no directional opacity) are densified as usual, while cone splats are added when we identify high-error view angles of a base splat. Periodically, each base splat counts, in every view, pixels with high L1 error that it covers, keeping track of view angles with many such pixels (\cref{fig:alt-spawn-cone}, left). If those views are much more concentrated than all views that see the splat, we add a cone splat copy of the base splat, with its axis $\mathbf{n}$ along the mean high-error direction and $k$ chosen for its cone to cover those views (\cref{fig:alt-spawn-cone}, right). The cone splats are added on top of the base splats rather than replacing them.

\subsection{Projection and Key Emission}
\label{sec:projkeys}
\papfig{Projection and key emission, ours (A) and original 3DGS (B)}{diagrams/fig_mech_projkeys_both}

\label{sec:keys-stage}
In forward projection, 3D splats are projected and generate tile-splat key pairs to be subsequently sorted by depth. This stage is primarily DRAM bandwidth-bound, having many Gaussian parameters to read and keys to write.

\parhead{Direct RGB\@.} By not using SH3, we already cut down many parameters that need to be moved. Yet additionally, by using flat RGB color as opposed to SH0, we can eliminate color computation in the forward projection stage, directly sending RGB to the forward rasterization stage when needed, thereby saving \qty{24}{\byte} of DRAM traffic per splat (\qty{12}{\byte} SH0 read, \qty{12}{\byte} RGB write).

\parhead{Projection and keys in 3DGS\@.} The original projection/keys stage repeats work, enumerating the Gaussian--tile intersections twice. The first computes and discards each splat's exact ellipse--tile spans to report only its total number of intersections, which is written to a global buffer. A prefix sum over the buffer obtains size M, which is copied to host for sizing the sort buffer. Only subsequently is each splat assigned a write offset. A second kernel re-reads every splat's projected state from memory, redoes the enumeration, and writes the recomputed Gaussian--tile pairs at the assigned offsets.

\parhead{Fused projection and keys.} We condense the two passes into one (\cref{fig:mech-projkeys-both})\vet{, as RoofGS~\cite{luo2026roofgs} does for rendering}. The sort buffer is pre-allocated at an ample capacity derived from peak historical iterations. Then projection proceeds, and a splat's ellipse--tile spans are calculated. 
Within a workgroup, each wave prefix-sums its lanes' pair counts locally, then does a single atomic addition on a global cursor to reserve a contiguous range in the pre-allocated sort buffer.
Finally, each lane derives its own write offset in the sort buffer from the prefix sum and workgroup's offset. This way, projection is only computed once, and key pairs are written once.

\parhead{Shortened keys.} VkSplat~\cite{vksplat} proposes shortened keys, which drops some bits from depth to shorten radix sort. The original key is $64$ bits, with $32$ bits of tile index, $32$ bits of depth. The shortened key is $32$ bits, with exact split depending on resolution (which determines number of tiles). With 5k resolution training images, $14$ bits are required for tiles and the remaining $18$ bits for depth, so $14$ of the $32$ depth bits are dropped and the key is half as long.

\subsection{Forward Rasterization}
\label{sec:fwd}
Each tile is rendered by one block of $256$ threads, one per pixel, in eight warps that each cover an $8{\times}4$ subtile, and the block loads the tile's depth-sorted splats into shared memory in batches of $256$. A splat seldom reaches the whole tile, so we adopt the subtile masks of VkSplat~\cite{vksplat}, which record which subtiles a splat's opacity-aware ellipse reaches, and, as in HiGS~\cite{higs}, each warp compacts the entries set for its subtile into a hit list in depth order before blending, so that it blends only the splats that reach it. With color staged in shared memory alongside the other parameters, the forward kernel is $1.22\times$ faster than with masks alone, with identical output.

\papfig{Strided pixel sampling}{diagrams/fig_strided_sampling}
\papfig{Loss, original with fused-ssim (left) and ours (right)}{diagrams/fig_mech_loss_pair_p1}
\parhead{Strided pixel sampling.}\label{sec:sampling}
We hypothesize that at high resolution, neighboring pixels give the splats they cover increasingly similar gradients, so a strided subset of the image can supervise competitively. Each step renders and supervises one pixel of every $2{\times}2$ block, at the same offset in every block, and the offset cycles through the four positions so that every pixel is supervised once every four steps (\cref{fig:sampling}). Unlike half-resolution training, each pixel is an actual captured pixel compared with the full-resolution ground truth, not an average of four. Turbo-GS~\cite{dhiman2024turbogs} introduced the idea as dilated rendering, with a random offset per iteration.

Hence, instead of a $16{\times}16$ tile size, we use a $32{\times}32$ tile size where only a quarter of the pixels are rasterized. An efficient implementation means this has equivalent performance as half-resolution rasterization. 

\subsection{Loss (L1 + SSIM)}
\label{sec:loss}

\label{sec:tile-ssim}

The $\mathcal{L}_1$ and SSIM terms of the training loss are evaluated over every
pixel of the rendered image, so cost scales with resolution independently
of scene complexity.

In the standard formulation, SSIM is split into the forward and backward SSIM passes. In the forward pass, SSIM is computed at every pixel over a sliding $11\times11$ 
\sg{Gaussian window, so each pixel draws on $121$ neighbors, which is computationally expensive}. A workgroup producing a $16\times16$ output tile must stage a $26\times26$ patch of both the rendered and the ground-truth image in shared memory, since the window reaches $5$ pixels beyond each edge of the tile ($5+16+5$).

In this formulation, backward pass furthermore requires intermediate derivatives from forward pass that likewise extends beyond its tile ($5+16+5$).
Hence, forward pass must first write these intermediate derivatives to global memory since backward needs derivatives from beyond the tile. The loss thus costs two dispatches and much read/write.

We replace the sliding window with a non-overlapping one \gm{(\cref{fig:mech-loss-pair})}. Taking the window size equal to the stride, ie. the $16\times16$ render tile, SSIM is evaluated exactly once per tile from only the pixels of that tile, weighted uniformly. Each pixel then contributes to exactly one statistic. There are two advantages.

First, no pixel outside the tile is required to be loaded to shared memory. Second, since the tile score $M$ depends on a pixel only through those moments, its gradient is available in closed form. No second convolution is needed (no separate forward and backward SSIM passes), and forward and backward fuse into a single dispatch in which no intermediate derivatives reach global memory. The actual SSIM value is never calculated or stored; we immediately calculate the relevant gradients for training. 

\papfig{Backward rasterization}{diagrams/fig_bwd_backward}
Lastly, since a fixed tile partition would place the window boundaries at the same pixels at every iteration, we offset the origin of the tile grid differently each iteration. 

Altogether, the loss stage costs \stLossStdLossX$\times$ less than with \texttt{fused-ssim}~\cite{mallick2024taming,fusedssim} (\stOursLossMs{} against \stLossStdLossMs\,ms per step) and a training step \stLossStdStepX$\times$ less (\cref{tab:stage_times}), with quality unchanged (PSNR \stOursPsnr{} against \stLossStdPsnr\,dB, SSIM \stOursSsim{} against \stLossStdSsim, mean over \stScenes{} scenes on an RTX~3070).


\subsection{Backward Rasterization}
\label{sec:bwd}

\label{sec:backward}
\parhead{Background.}
In backward rasterization, the two main families of algorithms are per-pixel and per-splat.

In per-pixel algorithms, threads map to pixels, and time steps through splats.
For example, a $16{\times}16$ tile of $256$ pixels is handled by $256$ threads, one thread handling one pixel. In every step, all $256$ threads compute gradients for a single splat, warp sums the gradients and sends an atomic increment for each gradient. Splats are processed every step till completion (\cref{fig:bwd}a).

In per-splat algorithms, threads map to splats, and time steps through pixels.
For example, $32$ splats are handled by $32$ threads, each thread handles one splat. In every step, each splat computes its gradient from one pixel. After $256$ steps, it finishes computing its gradient from a $16{\times}16$ tile ($256$ pixels), and sends an atomic increment for each of its gradients (\cref{fig:bwd}b).

\vet{Per-splat algorithms also proceed in a staircase-like manner, because of pixel state. Keeping a separate pixel state for every splat is too expensive (it would scale with pixels $\times$ splats), especially in shared memory (the L1 cache), so only one state is kept per pixel. A pixel must therefore meet its splats in order: a splat can process a pixel only after the previous splat is done with it. At timestep 0, thread 0 (splat 0) processes pixel 0, then hands that pixel's state to thread 1 (splat 1). At timestep 1, splat 0 processes pixel 1 while splat 1 processes pixel~0, and so on, each splat one pixel behind the one before it.}

\parhead{Note on checkpoints.}
In Taming-3DGS~\cite{mallick2024taming}, a per-splat algorithm, checkpoints are used where pixel state is snapshotted (during the forward pass) in global memory every $32$ splats, \ie the pixel state scales at pixels $\times$ splats$/32$ in global memory, but the pixel state loaded to L1 is still only one copy per $32$ splats. This is so that they can launch more jobs, but it does not make an individual job faster. On the contrary, this incurs overhead during the forward pass. It only benefits when the backward pass does not saturate the GPU with jobs, which does not happen at full resolution.

\parhead{Per-chunk traversal.}
We first describe our proposed algorithm, then offer some intuition on how to interpret it.
For every $16{\times}16$ tile $= 256$ pixels (8 chunks of $32$), we launch $128$ threads (4 warps of $32$). Each warp maps to a splat; at every timestep, a warp goes through a chunk, its $32$ threads processing the $32$ pixels of the chunk at once. Then in the next timestep, it processes another chunk. After 8 chunks (8 timesteps), a warp is done with a splat, sums the splat's gradients over its $32$ lanes once, and starts work on the splat that is 4 splats away from the given splat (since 3 other warps are working on the next 3 splats). This continues until the 4 warps have walked all of the tile's splats (\cref{fig:bwd}c).

The simplest intuition is that this is like per-splat, except instead of a thread, a warp maps to a splat. Instead of stepping one pixel at a time, we step through $32$ pixels (a chunk) at a time.

%
\parhead{\FpTrav{}.}
A splat seldom covers all of its tile (\cref{fig:bwd}d). Before walking, a splat tests its
footprint against the eight chunks, with the ellipse--box test the forward
uses for its subtile masks, and walks only the chunks it reaches.
With the reached chunks known, a splat waits only for the earlier splats that
reach the same chunk (\cref{fig:bwd}d). Each chunk counts the splats that have left it, and a splat enters a chunk once that count equals the number of earlier splats that reach it, so splats with disjoint footprints run side by side \vet{(\bwdAblStepSkipWait{} measured for both together)}.

\parhead{Arithmetic optimizations.}
\label{sec:bwd-arith}
\vet{We trim what each (splat, pixel) pair computes and what each pixel keeps.
These are pure performance optimizations.

\emph{Less state per pixel.} A pixel's accumulated color matters to $\partial L/\partial\alpha$ only through
its dot product with the pixel's color gradient $\mathbf v$. By keeping just $a$ instead of three color channels, A pixel's entry in shared memory shrinks from 36 to 24 bytes. It starts from minus the forward's final color, dotted with $\mathbf v$, and is updated by each pair:
\begin{equation}
  a \;\leftarrow\; a + \alpha T\,(\mathbf{rgb}\cdot\mathbf v),
  \qquad
  \frac{\partial L}{\partial\alpha} \;=\; T\,(\mathbf{rgb}\cdot\mathbf v) + \frac{a}{1-\alpha}.
  \label{eq:lean-dalpha}
\end{equation}

\emph{Splat constants applied once.} The conic and opacity are the same at every pixel of a splat. Instead of
multiplying them in at every pair, a pair adds to six running sums of $t = \partial L/\partial\alpha\cdot G$
weighted by the pixel's offset from the splat center. The splat only turns the sums into its gradients once, when it is done.

\emph{One exponential.} The conic is scaled once per splat by $-\tfrac12\log_2 e$, so each pair evaluates $G$ with a
single \texttt{exp2} and no extra multiply. The forward uses the same expression, so both passes keep the same pairs.}


\subsection{Backward Projection and Adam}
\label{sec:adam}

\papfig{Projection backward and Adam ablation}{diagrams/table_projadam_ablation_compact}

\label{sec:details}
In projection backward, the 2D gradients obtained prior are now backward projected to 3D gradients, applied to the splats, with Adam moments updated. This stage is also DRAM bandwidth bound with many existing optimizations available.

In gsplat~\cite{ye2024gsplat}, sparse Adam is offered, where unseen splats for that training view do not have Adam moments loaded or updated. Because only 33\% of splats on average are visible in each view (across 15 scenes, counting splats culled by their cone), this represents huge savings. We can check if a splat is visible via \#tiles, which is the number of tiles it covers, previously calculated in forward projection.
In efficient implementations like VkSplat~\cite{vksplat} and LichtFeld~\cite{lichtfeld}, the projection backward and Adam kernels are also fused; this avoids writing intermediate calculations and gradients to a buffer before Adam is applied.
Lastly, LichtFeld also uses compact moments, which reduces the VRAM usage of Adam moments but, more importantly, the DRAM bandwidth usage of this stage.

We merely adopt and combine these three ideas, specifically using bf16 for a compact moment (\cref{tab:projadam-ablation}). Coupled with our lightweight RGB and directional color, the stage scales well with splat count. \vet{\Cref{tab:projadam-ablation} summarises changes: a fused kernel with sparse Adam leaves an unseen splat costing 4 bytes read and 12 instructions (\paAblASpeed{} faster than gsplat). bf16 moments cut a visible splat's traffic from 0.7k to 0.5k bytes each way (\paAblBSpeed{}), and RGB with cone opacity cuts it to 0.2k, for \paAblCms{}\,ms for the whole stage against \paAblgsplatms{}\,ms.}

\subsection{Densification}
\label{sec:densify}
Densification follows FastGS~\cite{fastgs}, with a minor implementation change. FastGS computes a score during densification event by sampling and rendering training views, incurring a batch of extra forward passes. We instead accumulate the same statistic from renders throughout training, incurring no additional cost. When a training step ends, the composited image, per-pixel contributor counts, and ground truth are all still resident, making it an ideal time to take the metric with an additional pixel pass. We do so every 5 training steps.

\section{\ya{System Evaluation}}
\label{sec:results}


\begin{table*}[t]\centering
    \caption[Fastest on every GPU at comparable quality]{Fastest on every GPU at comparable quality.\ {\normalfont
   SoTA vs. {\name}, trained at full res. for 30k steps on Mip-NeRF~360 and Tanks\&Temples~4K on three consumer GPUs and an RTX~4090. \rankbox{rankone}{Red}, \rankbox{ranktwo}{orange}, 
    \rankbox{rankthree}{yellow}: first, second, third.}}
    \footnotesize
    \setlength{\tabcolsep}{1.8pt}
    \begin{tabular*}{\textwidth}{@{\extracolsep{\fill}}l rrrrr @{\hskip2pt\vrule width0.4pt\hskip2pt} rrrr @{\hskip3pt\vrule width0.9pt\hskip3pt} rrrrr @{\hskip2pt\vrule width0.4pt\hskip2pt} rrrr@{}}
        \toprule
         & \multicolumn{9}{c}{Mip-NeRF 360 (7 scenes)} & \multicolumn{9}{c}{Tanks \& Temples 4K (8 scenes)} \\
        \cmidrule(r{2pt}){2-10}\cmidrule(l{2pt}){11-19}
        Method & PSNR$\uparrow$ & SSIM$\uparrow$ & LPIPS$\downarrow$ & Splats & MB & RTX3060 & RTX3070 & RTX3090 & RTX4090 & PSNR$\uparrow$ & SSIM$\uparrow$ & LPIPS$\downarrow$ & Splats & MB & RTX3060 & RTX3070 & RTX3090 & RTX4090 \\
        \midrule
        Mip-Splatting~\cite{yu2024mipsplatting} & 28.83 & \cellcolor{rankone}0.866 & \cellcolor{rankone}0.234 & 3.59M & 906 & -- & -- & -- & 103m 43s & 23.34 & \cellcolor{rankthree}0.813 & \cellcolor{ranktwo}0.251 & 2.71M & 682 & -- & -- & -- & 87m 23s \\
        Speedy-Splat~\cite{hanson2025speedysplat} & 27.57 & 0.824 & 0.366 & 0.20M & 51 & -- & -- & 88m 59s & \phantom{0}70m \phantom{0}7s & 22.92 & 0.778 & 0.371 & 0.16M & 39 & -- & -- & 79m 20s & 64m \phantom{0}0s \\
        Taming-3DGS~\cite{mallick2024taming} & 28.26 & 0.840 & 0.328 & 0.62M & 153 & -- & -- & 53m 15s & \phantom{0}25m \phantom{0}9s & 23.62 & 0.787 & 0.349 & 0.33M & 82 & -- & -- & 39m 33s & 19m 14s \\
        DashGaussian~\cite{chen2025dashgaussian} & 28.57 & 0.853 & 0.272 & 2.14M & 531 & -- & -- & 42m 20s & \phantom{0}19m 50s & 24.03 & 0.807 & 0.283 & 1.60M & 398 & -- & -- & 30m \phantom{0}0s & 15m 13s \\
        Faster-GS~\cite{hahlbohm2026fastergs} & 28.45 & 0.851 & 0.293 & 1.92M & 453 & 48m \phantom{0}8s & 33m \phantom{0}3s & 19m 50s & \phantom{0}16m 31s & 23.49 & 0.807 & 0.280 & 1.59M & 375 & 50m \phantom{0}4s & 30m 39s & 19m 30s & 11m 58s \\
        VkSplat~\cite{vksplat} & 28.39 & 0.850 & 0.297 & 2.35M & 556 & 40m 59s & 27m \phantom{0}3s & 16m 20s & \phantom{0}10m 32s & 23.14 & 0.803 & 0.291 & 1.79M & 423 & 37m 44s & 23m 25s & 13m 51s & \phantom{0}8m \phantom{0}7s \\
        LiteGS~\cite{liao2026litegs} & \cellcolor{rankone}29.30 & \cellcolor{ranktwo}0.863 & \cellcolor{rankthree}0.256 & 0.99M & 246 & \cellcolor{ranktwo}34m \phantom{0}8s & \cellcolor{ranktwo}23m 54s & 15m 38s & \cellcolor{rankthree}\phantom{00}8m 17s & \cellcolor{rankone}24.84 & \cellcolor{ranktwo}0.819 & \cellcolor{rankthree}0.252 & 0.99M & 246 & \cellcolor{ranktwo}32m \phantom{0}2s & \cellcolor{rankthree}22m 23s & 14m 54s & \phantom{0}8m \phantom{0}5s \\
        LichtFeld~\cite{lichtfeld} & \cellcolor{ranktwo}29.14 & \cellcolor{ranktwo}0.863 & \cellcolor{ranktwo}0.252 & 1.00M & 248 & 38m 35s & 28m 30s & \cellcolor{rankthree}14m 52s & \cellcolor{ranktwo}\phantom{00}8m \phantom{0}7s & \cellcolor{rankthree}24.21 & \cellcolor{rankone}0.821 & \cellcolor{rankone}0.236 & 1.00M & 248 & 38m 46s & 26m 16s & \cellcolor{rankthree}13m 50s & \cellcolor{rankthree}\phantom{0}7m 23s \\
        FastGS~\cite{fastgs} & 28.36 & 0.847 & 0.299 & 0.54M & 134 & \cellcolor{rankthree}37m 17s & \cellcolor{rankthree}25m \phantom{0}1s & \cellcolor{ranktwo}14m 11s & \phantom{00}8m 22s & 23.31 & 0.797 & 0.308 & 0.43M & 107 & \cellcolor{rankthree}34m \phantom{0}8s & \cellcolor{ranktwo}21m 12s & \cellcolor{ranktwo}12m \phantom{0}0s & \cellcolor{ranktwo}\phantom{0}6m 57s \\
        \midrule
        \textbf{{\name} (Ours)} & \cellcolor{rankthree}28.96 & \cellcolor{rankthree}0.855 & 0.263 & 1.57M & 95 & \cellcolor{rankone}\phantom{0}6m 25s & \cellcolor{rankone}\phantom{0}4m 24s & \cellcolor{rankone}\phantom{0}2m 56s & \cellcolor{rankone}\phantom{00}1m 29s & \cellcolor{ranktwo}24.35 & \cellcolor{rankthree}0.813 & 0.258 & 1.71M & 112 & \cellcolor{rankone}\phantom{0}6m 37s & \cellcolor{rankone}\phantom{0}4m 36s & \cellcolor{rankone}\phantom{0}3m \phantom{0}6s & \cellcolor{rankone}\phantom{0}1m 38s \\
        \bottomrule
    \end{tabular*}
    \label{tab:main_comparison}
\end{table*}

%
%

\subsection{\ya{Experimental} Setup}
\label{sec:results:setup}

\parhead{\ya{Datasets.}} Fifteen scenes are used, including seven from Mip-NeRF~360~\cite{barron2022mipnerf360} at full resolution and eight from Tanks\&Temples~\cite{knapitsch2017tanks} re-extracted at 4K resolution from source video.

\parhead{\ya{Comparisons.}} We evaluate training on four different consumer-grade GPU models, NVIDIA GeForce RTX 3060, 3070, 3090, 4090 against nine trainers using their official code. 
Taming-3DGS~\cite{mallick2024taming}, DashGaussian~\cite{chen2025dashgaussian}, Speedy-Splat~\cite{hanson2025speedysplat}, FastGS~\cite{fastgs} and Faster-GS~\cite{hahlbohm2026fastergs} accelerate
 the original 3DGS training, mainly through densification, pruning or schedule of training resolution.
VkSplat~\cite{vksplat}, LiteGS~\cite{liao2026litegs} and LichtFeld~\cite{lichtfeld} are trainers re-engineered for speed, closest to ours.
Mip-Splatting~\cite{yu2024mipsplatting} is included as a quality reference that does not target speed.
Each method runs with its default settings, except Taming-3DGS in its budget mode and LichtFeld in its MRNF benchmark configuration (1M splats, exposure correction and depth and normal generation off).

\parhead{\ya{Metrics.}} All evaluations use full-resolution. SSIM~\cite{wang2004image} is taken with the reference formulation (over every pixel with an $11{\times}11$
window) and not from our discounted version.
LPIPS is LPIPS-Alex~\cite{zhang2018unreasonable} at full-resolution.
Every eighth image of a scene is held out for testing.

Peak host memory and peak VRAM are reported for our method in \cref{sec:results:main}; they were not recorded for the other methods.
Quality is averaged over runs, then GPUs, then scenes; time is averaged over the runs of a scene on one GPU model, then over scenes.


\subsection{\ya{Overall Performance}}
\label{sec:results:main}
\newcommand{\VramMipMean}{0.96}
\newcommand{\VramMipMax}{1.76}
\newcommand{\VramMipMaxScene}{bicycle}
\newcommand{\VramTntMean}{1.11}
\newcommand{\VramTntMax}{1.46}
\newcommand{\VramTntMaxScene}{ignatius}
\newcommand{\VramAllMean}{1.04}
\newcommand{\VramAllMax}{1.76}
\newcommand{\VramAllMaxScene}{bicycle}
\newcommand{\VramAllN}{15}
\newcommand{\HostAllMean}{9.89}
\newcommand{\HostAllMax}{13.5}
\newcommand{\HostAllMaxScene}{garden}
\newcommand{\HostAllN}{15}
\newcommand{\ConeAllPct}{48}
\newcommand{\BaseAllPct}{52}

\ya{We evaluate \method{} as a complete system as shown in \Cref{tab:main_comparison}. Several observations can be made.}

\parhead{Training time.} At 30k steps, our \method{} is the fastest of \cmpMethods{} methods on both datasets and all four GPUs (\cref{tab:main_comparison}), \cmpGainShortMin--\cmpGainShortMax$\times$ faster than the next fastest, and it trains a scene in under 7 minutes on every GPU and in under 2 minutes on the RTX~4090. On the \cmpEntryCard{}, \cmpEntryNoTime{} of the \cmpEntryRivals{} other methods cannot be trained in practice, and the fastest of the rest, LiteGS, needs 34m\,8s per Mip-NeRF~360 scene against \cmpOursShortEntryTime{} for \method{}. On both datasets, \method{} trains faster on the \cmpEntryCard{} than any other method does on the RTX~4090: \cmpOursShortEntryTime{} against 8m\,7s for LichtFeld on Mip-NeRF~360, and 6m\,37s against 6m\,57s for FastGS on Tanks\&Temples~4K.

\parhead{Quality.} \ya{ The speedup comes with little loss in quality. At \num{30}k steps, \method{} ranks third in PSNR on Mip-NeRF~360 (28.96\,dB) and second on Tanks\&Temples~4K (24.35\,dB), and it outperforms FastGS, the fastest prior trainer on half of the GPU--dataset pairs, on every quality metric of both datasets.}

\parhead{Host memory usage.} The full resolution training images stay in host memory and are transferred (triple buffered) to VRAM when needed for training. Mean peak host memory was \HostAllMean{}\,GB (maximum \HostAllMax{}\,GB on \HostAllMaxScene) over all \HostAllN{} scenes.

\parhead{VRAM usage.} Mean peak VRAM was \VramAllMean{}\,GB (maximum \VramAllMax{}\,GB on \VramAllMaxScene) over all \VramAllN{} scenes. There is flexibility since training images can stay in VRAM if host memory turns out to be limiting.

\parhead{File size.} While our splat counts are not the lowest, the file size works out to be rather small as each splat has far fewer parameters: \SI{72}{\byte} per cone splat and \SI{56}{\byte} per base splat. On average over all scenes, \ConeAllPct\% of splats are cone splats and \BaseAllPct\% base splats.


\parhead{Culling.} Over our \coneCullScenes{} scenes, the cone culls on average \coneCullOfVisible\% of the splats a view would otherwise draw (\coneCullOfVisibleMin--\coneCullOfVisibleMax\% per scene), \coneCullOfAll\% of all splats.
A cone splat is more often inside a given view than a base splat (\coneInViewPct\% against \baseInViewPct\%), and the cone culls about half (\coneCullOfConeInView\%) of the cone splats in view. The culled splats cover more tiles than the average splat, so culling removes \coneCullPairs\% of a view's Gaussian--tile pairs, which then skip the sort, both rasterization passes and the Adam step.

\subsection{Ablations}
\label{sec:ablation}

\begin{table*}[t]\centering
\caption[Ablations]{Ablations.\ {\normalfont
Training time by stage on an RTX~3070 (s, with the share of the total). Each ``w/o'' row puts one stage back to its standard implementation and keeps all others; \rankbox{rankone}{highlighted cells} mark the stages it targets. All Changes removes every single change.}}
\footnotesize
\setlength{\tabcolsep}{3pt}
\begin{tabular*}{\textwidth}{@{\extracolsep{\fill}}l | rrrr | c | c c c c c c c@{}}
\toprule
Method & PSNR$\uparrow$ & SSIM$\uparrow$ & LPIPS$\downarrow$ & MB & Total & Proj.\ Fwd & Sort & Raster Fwd & Loss & Raster Bwd & Proj.\ Bwd + Adam & Other \\
\midrule
{\name} (Ours) & 26.48 & 0.832 & 0.287 & 118 & \makebox[2.05em][r]{269}\hspace{1.5pt}\makebox[3.1em][l]{\color{black!50}(1$\times$)} & \makebox[2.05em][r]{12}\hspace{1.5pt}\makebox[2.6em][l]{{\color{black!50}(4\%)}} & \makebox[2.05em][r]{19}\hspace{1.5pt}\makebox[2.6em][l]{{\color{black!50}(7\%)}} & \makebox[2.05em][r]{46}\hspace{1.5pt}\makebox[2.6em][l]{{\color{black!50}(17\%)}} & \makebox[2.05em][r]{7}\hspace{1.5pt}\makebox[2.6em][l]{{\color{black!50}(3\%)}} & \makebox[2.05em][r]{147}\hspace{1.5pt}\makebox[2.6em][l]{{\color{black!50}(55\%)}} & \makebox[2.05em][r]{17}\hspace{1.5pt}\makebox[2.6em][l]{{\color{black!50}(6\%)}} & \makebox[2.05em][r]{21}\hspace{1.5pt}\makebox[2.6em][l]{{\color{black!50}(8\%)}} \\
\midrule
w/o Fused Keys, Sort & 26.47 & 0.832 & 0.287 & 118 & \makebox[2.05em][r]{300}\hspace{1.5pt}\makebox[3.1em][l]{\color{black!50}(1.11$\times$)} & \cellcolor{rankone}\makebox[2.05em][r]{19}\hspace{1.5pt}\makebox[2.6em][l]{{\color{black!50}(6\%)}} & \cellcolor{rankone}\makebox[2.05em][r]{44}\hspace{1.5pt}\makebox[2.6em][l]{{\color{black!50}(15\%)}} & \makebox[2.05em][r]{46}\hspace{1.5pt}\makebox[2.6em][l]{{\color{black!50}(15\%)}} & \makebox[2.05em][r]{7}\hspace{1.5pt}\makebox[2.6em][l]{{\color{black!50}(2\%)}} & \makebox[2.05em][r]{146}\hspace{1.5pt}\makebox[2.6em][l]{{\color{black!50}(49\%)}} & \makebox[2.05em][r]{17}\hspace{1.5pt}\makebox[2.6em][l]{{\color{black!50}(6\%)}} & \makebox[2.05em][r]{20}\hspace{1.5pt}\makebox[2.6em][l]{{\color{black!50}(7\%)}} \\
w/o Raster Fwd, Sampl. & 26.50 & 0.832 & 0.286 & 117 & \makebox[2.05em][r]{904}\hspace{1.5pt}\makebox[3.1em][l]{\color{black!50}(3.35$\times$)} & \makebox[2.05em][r]{12}\hspace{1.5pt}\makebox[2.6em][l]{{\color{black!50}(1\%)}} & \makebox[2.05em][r]{19}\hspace{1.5pt}\makebox[2.6em][l]{{\color{black!50}(2\%)}} & \cellcolor{rankone}\makebox[2.05em][r]{232}\hspace{1.5pt}\makebox[2.6em][l]{{\color{black!50}(26\%)}} & \makebox[2.05em][r]{28}\hspace{1.5pt}\makebox[2.6em][l]{{\color{black!50}(3\%)}} & \makebox[2.05em][r]{575}\hspace{1.5pt}\makebox[2.6em][l]{{\color{black!50}(64\%)}} & \makebox[2.05em][r]{17}\hspace{1.5pt}\makebox[2.6em][l]{{\color{black!50}(2\%)}} & \makebox[2.05em][r]{20}\hspace{1.5pt}\makebox[2.6em][l]{{\color{black!50}(2\%)}} \\
w/o Tile SSIM & 26.34 & 0.833 & 0.290 & 112 & \makebox[2.05em][r]{311}\hspace{1.5pt}\makebox[3.1em][l]{\color{black!50}(1.16$\times$)} & \makebox[2.05em][r]{11}\hspace{1.5pt}\makebox[2.6em][l]{{\color{black!50}(4\%)}} & \makebox[2.05em][r]{19}\hspace{1.5pt}\makebox[2.6em][l]{{\color{black!50}(6\%)}} & \makebox[2.05em][r]{45}\hspace{1.5pt}\makebox[2.6em][l]{{\color{black!50}(14\%)}} & \cellcolor{rankone}\makebox[2.05em][r]{60}\hspace{1.5pt}\makebox[2.6em][l]{{\color{black!50}(19\%)}} & \makebox[2.05em][r]{140}\hspace{1.5pt}\makebox[2.6em][l]{{\color{black!50}(45\%)}} & \makebox[2.05em][r]{17}\hspace{1.5pt}\makebox[2.6em][l]{{\color{black!50}(5\%)}} & \makebox[2.05em][r]{20}\hspace{1.5pt}\makebox[2.6em][l]{{\color{black!50}(7\%)}} \\
w/o Raster Bwd & 26.49 & 0.832 & 0.287 & 118 & \makebox[2.05em][r]{651}\hspace{1.5pt}\makebox[3.1em][l]{\color{black!50}(2.42$\times$)} & \makebox[2.05em][r]{12}\hspace{1.5pt}\makebox[2.6em][l]{{\color{black!50}(2\%)}} & \makebox[2.05em][r]{20}\hspace{1.5pt}\makebox[2.6em][l]{{\color{black!50}(3\%)}} & \makebox[2.05em][r]{46}\hspace{1.5pt}\makebox[2.6em][l]{{\color{black!50}(7\%)}} & \makebox[2.05em][r]{7}\hspace{1.5pt}\makebox[2.6em][l]{{\color{black!50}(1\%)}} & \cellcolor{rankone}\makebox[2.05em][r]{530}\hspace{1.5pt}\makebox[2.6em][l]{{\color{black!50}(81\%)}} & \makebox[2.05em][r]{17}\hspace{1.5pt}\makebox[2.6em][l]{{\color{black!50}(3\%)}} & \makebox[2.05em][r]{20}\hspace{1.5pt}\makebox[2.6em][l]{{\color{black!50}(3\%)}} \\
w/o Fused Adam & 26.49 & 0.832 & 0.287 & 118 & \makebox[2.05em][r]{315}\hspace{1.5pt}\makebox[3.1em][l]{\color{black!50}(1.17$\times$)} & \makebox[2.05em][r]{12}\hspace{1.5pt}\makebox[2.6em][l]{{\color{black!50}(4\%)}} & \makebox[2.05em][r]{19}\hspace{1.5pt}\makebox[2.6em][l]{{\color{black!50}(6\%)}} & \makebox[2.05em][r]{46}\hspace{1.5pt}\makebox[2.6em][l]{{\color{black!50}(15\%)}} & \makebox[2.05em][r]{7}\hspace{1.5pt}\makebox[2.6em][l]{{\color{black!50}(2\%)}} & \makebox[2.05em][r]{146}\hspace{1.5pt}\makebox[2.6em][l]{{\color{black!50}(46\%)}} & \cellcolor{rankone}\makebox[2.05em][r]{63}\hspace{1.5pt}\makebox[2.6em][l]{{\color{black!50}(20\%)}} & \makebox[2.05em][r]{21}\hspace{1.5pt}\makebox[2.6em][l]{{\color{black!50}(7\%)}} \\
w/o RGB + Cone & 25.99 & 0.829 & 0.293 & 278 & \makebox[2.05em][r]{263}\hspace{1.5pt}\makebox[3.1em][l]{\color{black!50}(0.98$\times$)} & \cellcolor{rankone}\makebox[2.05em][r]{17}\hspace{1.5pt}\makebox[2.6em][l]{{\color{black!50}(6\%)}} & \makebox[2.05em][r]{17}\hspace{1.5pt}\makebox[2.6em][l]{{\color{black!50}(7\%)}} & \makebox[2.05em][r]{39}\hspace{1.5pt}\makebox[2.6em][l]{{\color{black!50}(15\%)}} & \makebox[2.05em][r]{7}\hspace{1.5pt}\makebox[2.6em][l]{{\color{black!50}(3\%)}} & \makebox[2.05em][r]{122}\hspace{1.5pt}\makebox[2.6em][l]{{\color{black!50}(47\%)}} & \cellcolor{rankone}\makebox[2.05em][r]{38}\hspace{1.5pt}\makebox[2.6em][l]{{\color{black!50}(14\%)}} & \makebox[2.05em][r]{24}\hspace{1.5pt}\makebox[2.6em][l]{{\color{black!50}(9\%)}} \\
\midrule
w/o All Changes & 25.84 & 0.830 & 0.297 & 254 & \makebox[2.05em][r]{2326}\hspace{1.5pt}\makebox[3.1em][l]{\color{black!50}(8.63$\times$)} & \makebox[2.05em][r]{31}\hspace{1.5pt}\makebox[2.6em][l]{{\color{black!50}(1\%)}} & \makebox[2.05em][r]{37}\hspace{1.5pt}\makebox[2.6em][l]{{\color{black!50}(2\%)}} & \makebox[2.05em][r]{182}\hspace{1.5pt}\makebox[2.6em][l]{{\color{black!50}(8\%)}} & \makebox[2.05em][r]{235}\hspace{1.5pt}\makebox[2.6em][l]{{\color{black!50}(10\%)}} & \makebox[2.05em][r]{1682}\hspace{1.5pt}\makebox[2.6em][l]{{\color{black!50}(72\%)}} & \makebox[2.05em][r]{136}\hspace{1.5pt}\makebox[2.6em][l]{{\color{black!50}(6\%)}} & \makebox[2.05em][r]{23}\hspace{1.5pt}\makebox[2.6em][l]{{\color{black!50}(1\%)}} \\
\bottomrule
\end{tabular*}
\label{tab:stage_times}
\end{table*}


In \cref{tab:stage_times}, we separately remove the improvements of each stage in \cref{sec:method} from the full system and measure the per-stage training times on an RTX~3070. The impact of each change is targetted and indeed impacts its own stage.

\newcommand{\sdKeysTwopassProjMs}{0.74}
\newcommand{\sdKeysTwopassProjSlow}{1.59$\times$}
\newcommand{\sdKeysTwopassProjDvram}{116}
\newcommand{\sdKeysTwopassProjVram}{1{,}208}
\newcommand{\sdKeysTwopassProjKeys}{1.00}
\newcommand{\sdKeysTwopassProjInstr}{68}
\newcommand{\sdKeysTwopassProjDram}{286}
\newcommand{\sdKeysTwopassProjKernels}{5}
\newcommand{\sdKeysTwopassProjRegs}{40}
\newcommand{\sdKeysTwopassProjThreads}{128}
\newcommand{\sdKeysTwopassProjSmem}{0}
\newcommand{\sdKeysTwopassProjOcc}{90\%}
\newcommand{\sdKeysTwopassProjOccTh}{100\%}
\newcommand{\sdKeysTwopassProjIssue}{32\%}
\newcommand{\sdKeysTwopassProjSpeed}{1.00$\times$}
\newcommand{\sdKeysFusedNorejectProjMs}{0.60}
\newcommand{\sdKeysFusedNorejectProjSlow}{1.29$\times$}
\newcommand{\sdKeysFusedNorejectProjDvram}{0}
\newcommand{\sdKeysFusedNorejectProjVram}{1{,}092}
\newcommand{\sdKeysFusedNorejectProjKeys}{1.00}
\newcommand{\sdKeysFusedNorejectProjInstr}{75}
\newcommand{\sdKeysFusedNorejectProjDram}{221}
\newcommand{\sdKeysFusedNorejectProjKernels}{1}
\newcommand{\sdKeysFusedNorejectProjRegs}{40}
\newcommand{\sdKeysFusedNorejectProjThreads}{128}
\newcommand{\sdKeysFusedNorejectProjSmem}{24}
\newcommand{\sdKeysFusedNorejectProjOcc}{94\%}
\newcommand{\sdKeysFusedNorejectProjOccTh}{100\%}
\newcommand{\sdKeysFusedNorejectProjIssue}{40\%}
\newcommand{\sdKeysFusedNorejectProjSpeed}{1.24$\times$}
\newcommand{\sdKeysFusedProjMs}{0.56}
\newcommand{\sdKeysFusedProjSlow}{1.21$\times$}
\newcommand{\sdKeysFusedProjDvram}{0}
\newcommand{\sdKeysFusedProjVram}{1{,}092}
\newcommand{\sdKeysFusedProjKeys}{1.00}
\newcommand{\sdKeysFusedProjInstr}{71}
\newcommand{\sdKeysFusedProjDram}{175}
\newcommand{\sdKeysFusedProjKernels}{1}
\newcommand{\sdKeysFusedProjRegs}{40}
\newcommand{\sdKeysFusedProjThreads}{128}
\newcommand{\sdKeysFusedProjSmem}{24}
\newcommand{\sdKeysFusedProjOcc}{94\%}
\newcommand{\sdKeysFusedProjOccTh}{100\%}
\newcommand{\sdKeysFusedProjIssue}{41\%}
\newcommand{\sdKeysFusedProjSpeed}{1.31$\times$}
\newcommand{\sdKeysCursorNorejectProjMs}{0.53}
\newcommand{\sdKeysCursorNorejectProjSlow}{1.15$\times$}
\newcommand{\sdKeysCursorNorejectProjDvram}{0}
\newcommand{\sdKeysCursorNorejectProjVram}{1{,}092}
\newcommand{\sdKeysCursorNorejectProjKeys}{1.00}
\newcommand{\sdKeysCursorNorejectProjInstr}{68}
\newcommand{\sdKeysCursorNorejectProjDram}{221}
\newcommand{\sdKeysCursorNorejectProjKernels}{2}
\newcommand{\sdKeysCursorNorejectProjRegs}{40}
\newcommand{\sdKeysCursorNorejectProjThreads}{128}
\newcommand{\sdKeysCursorNorejectProjSmem}{0}
\newcommand{\sdKeysCursorNorejectProjOcc}{79\%}
\newcommand{\sdKeysCursorNorejectProjOccTh}{100\%}
\newcommand{\sdKeysCursorNorejectProjIssue}{43\%}
\newcommand{\sdKeysCursorNorejectProjSpeed}{1.38$\times$}
\newcommand{\sdOursProjMs}{0.46}
\newcommand{\sdOursProjSlow}{1.00$\times$}
\newcommand{\sdOursProjDvram}{0}
\newcommand{\sdOursProjVram}{1{,}092}
\newcommand{\sdOursProjKeys}{1.00}
\newcommand{\sdOursProjInstr}{63}
\newcommand{\sdOursProjDram}{175}
\newcommand{\sdOursProjKernels}{2}
\newcommand{\sdOursProjRegs}{40}
\newcommand{\sdOursProjThreads}{128}
\newcommand{\sdOursProjSmem}{0}
\newcommand{\sdOursProjOcc}{82\%}
\newcommand{\sdOursProjOccTh}{100\%}
\newcommand{\sdOursProjIssue}{48\%}
\newcommand{\sdOursProjSpeed}{1.59$\times$}
\newcommand{\sdNodirectProjMs}{0.51}
\newcommand{\sdNodirectProjSlow}{1.11$\times$}
\newcommand{\sdNodirectProjDvram}{80}
\newcommand{\sdNodirectProjVram}{1{,}172}
\newcommand{\sdNodirectProjKeys}{0.99}
\newcommand{\sdNodirectProjInstr}{65}
\newcommand{\sdNodirectProjDram}{210}
\newcommand{\sdNodirectProjKernels}{2}
\newcommand{\sdNodirectProjRegs}{40}
\newcommand{\sdNodirectProjThreads}{128}
\newcommand{\sdNodirectProjSmem}{0}
\newcommand{\sdNodirectProjOcc}{82\%}
\newcommand{\sdNodirectProjOccTh}{100\%}
\newcommand{\sdNodirectProjIssue}{43\%}
\newcommand{\sdNodirectProjSpeed}{1.43$\times$}
\newcommand{\sdKeysTwopassShZeroProjMs}{--}
\newcommand{\sdKeysTwopassShZeroProjSlow}{--$\times$}
\newcommand{\sdKeysTwopassShZeroProjDvram}{--}
\newcommand{\sdKeysTwopassShZeroProjVram}{--}
\newcommand{\sdKeysTwopassShZeroProjKeys}{--}
\newcommand{\sdKeysTwopassShZeroProjInstr}{69}
\newcommand{\sdKeysTwopassShZeroProjDram}{339}
\newcommand{\sdKeysTwopassShZeroProjKernels}{5}
\newcommand{\sdKeysTwopassShZeroProjRegs}{48}
\newcommand{\sdKeysTwopassShZeroProjThreads}{128}
\newcommand{\sdKeysTwopassShZeroProjSmem}{0}
\newcommand{\sdKeysTwopassShZeroProjOcc}{73\%}
\newcommand{\sdKeysTwopassShZeroProjOccTh}{83\%}
\newcommand{\sdKeysTwopassShZeroProjIssue}{25\%}
\newcommand{\sdKeysCursorNorejectShZeroProjMs}{--}
\newcommand{\sdKeysCursorNorejectShZeroProjSlow}{--$\times$}
\newcommand{\sdKeysCursorNorejectShZeroProjDvram}{--}
\newcommand{\sdKeysCursorNorejectShZeroProjVram}{--}
\newcommand{\sdKeysCursorNorejectShZeroProjKeys}{--}
\newcommand{\sdKeysCursorNorejectShZeroProjInstr}{70}
\newcommand{\sdKeysCursorNorejectShZeroProjDram}{273}
\newcommand{\sdKeysCursorNorejectShZeroProjKernels}{2}
\newcommand{\sdKeysCursorNorejectShZeroProjRegs}{40}
\newcommand{\sdKeysCursorNorejectShZeroProjThreads}{128}
\newcommand{\sdKeysCursorNorejectShZeroProjSmem}{0}
\newcommand{\sdKeysCursorNorejectShZeroProjOcc}{79\%}
\newcommand{\sdKeysCursorNorejectShZeroProjOccTh}{100\%}
\newcommand{\sdKeysCursorNorejectShZeroProjIssue}{36\%}
\newcommand{\sdSortStdSortMs}{1.81}
\newcommand{\sdSortStdSortSlow}{2.22$\times$}
\newcommand{\sdSortStdSortDvram}{92}
\newcommand{\sdSortStdSortVram}{1{,}184}
\newcommand{\sdSortStdSortKeys}{1.00}
\newcommand{\sdSortStdSortInstr}{95}
\newcommand{\sdSortStdSortDram}{788}
\newcommand{\sdSortStdSortKernels}{9}
\newcommand{\sdSortStdSortRegs}{56}
\newcommand{\sdSortStdSortThreads}{384}
\newcommand{\sdSortStdSortSmem}{31{,}744}
\newcommand{\sdSortStdSortOcc}{73\%}
\newcommand{\sdSortStdSortOccTh}{75\%}
\newcommand{\sdSortStdSortIssue}{19\%}
\newcommand{\sdSortStdSortSpeed}{1.00$\times$}
\newcommand{\sdOursSortMs}{0.81}
\newcommand{\sdOursSortSlow}{1.00$\times$}
\newcommand{\sdOursSortDvram}{0}
\newcommand{\sdOursSortVram}{1{,}092}
\newcommand{\sdOursSortKeys}{1.00}
\newcommand{\sdOursSortInstr}{62}
\newcommand{\sdOursSortDram}{362}
\newcommand{\sdOursSortKernels}{7}
\newcommand{\sdOursSortRegs}{80}
\newcommand{\sdOursSortThreads}{384}
\newcommand{\sdOursSortSmem}{33{,}280}
\newcommand{\sdOursSortOcc}{48\%}
\newcommand{\sdOursSortOccTh}{50\%}
\newcommand{\sdOursSortIssue}{25\%}
\newcommand{\sdOursSortSpeed}{2.22$\times$}
\newcommand{\sdFwdNomaskFwdMs}{2.73}
\newcommand{\sdFwdNomaskFwdSlow}{1.41$\times$}
\newcommand{\sdFwdNomaskFwdDvram}{0}
\newcommand{\sdFwdNomaskFwdVram}{1{,}092}
\newcommand{\sdFwdNomaskFwdKeys}{1.00}
\newcommand{\sdFwdNomaskFwdInstr}{1{,}023}
\newcommand{\sdFwdNomaskFwdDram}{152}
\newcommand{\sdFwdNomaskFwdKernels}{1}
\newcommand{\sdFwdNomaskFwdRegs}{33}
\newcommand{\sdFwdNomaskFwdThreads}{256}
\newcommand{\sdFwdNomaskFwdSmem}{10{,}240}
\newcommand{\sdFwdNomaskFwdOcc}{96\%}
\newcommand{\sdFwdNomaskFwdOccTh}{100\%}
\newcommand{\sdFwdNomaskFwdIssue}{85\%}
\newcommand{\sdFwdNomaskFwdSpeed}{1.00$\times$}
\newcommand{\sdFwdMasksFwdMs}{2.38}
\newcommand{\sdFwdMasksFwdSlow}{1.23$\times$}
\newcommand{\sdFwdMasksFwdDvram}{0}
\newcommand{\sdFwdMasksFwdVram}{1{,}092}
\newcommand{\sdFwdMasksFwdKeys}{1.00}
\newcommand{\sdFwdMasksFwdInstr}{877}
\newcommand{\sdFwdMasksFwdDram}{152}
\newcommand{\sdFwdMasksFwdKernels}{1}
\newcommand{\sdFwdMasksFwdRegs}{40}
\newcommand{\sdFwdMasksFwdThreads}{256}
\newcommand{\sdFwdMasksFwdSmem}{10{,}240}
\newcommand{\sdFwdMasksFwdOcc}{95\%}
\newcommand{\sdFwdMasksFwdOccTh}{100\%}
\newcommand{\sdFwdMasksFwdIssue}{84\%}
\newcommand{\sdFwdMasksFwdSpeed}{1.15$\times$}
\newcommand{\sdFwdColorFwdMs}{2.27}
\newcommand{\sdFwdColorFwdSlow}{1.17$\times$}
\newcommand{\sdFwdColorFwdDvram}{0}
\newcommand{\sdFwdColorFwdVram}{1{,}092}
\newcommand{\sdFwdColorFwdKeys}{1.00}
\newcommand{\sdFwdColorFwdInstr}{859}
\newcommand{\sdFwdColorFwdDram}{152}
\newcommand{\sdFwdColorFwdKernels}{1}
\newcommand{\sdFwdColorFwdRegs}{39}
\newcommand{\sdFwdColorFwdThreads}{256}
\newcommand{\sdFwdColorFwdSmem}{11{,}264}
\newcommand{\sdFwdColorFwdOcc}{95\%}
\newcommand{\sdFwdColorFwdOccTh}{100\%}
\newcommand{\sdFwdColorFwdIssue}{86\%}
\newcommand{\sdFwdColorFwdSpeed}{1.20$\times$}
\newcommand{\sdFwdHitsFwdMs}{2.17}
\newcommand{\sdFwdHitsFwdSlow}{1.13$\times$}
\newcommand{\sdFwdHitsFwdDvram}{0}
\newcommand{\sdFwdHitsFwdVram}{1{,}092}
\newcommand{\sdFwdHitsFwdKeys}{1.00}
\newcommand{\sdFwdHitsFwdInstr}{711}
\newcommand{\sdFwdHitsFwdDram}{152}
\newcommand{\sdFwdHitsFwdKernels}{1}
\newcommand{\sdFwdHitsFwdRegs}{38}
\newcommand{\sdFwdHitsFwdThreads}{256}
\newcommand{\sdFwdHitsFwdSmem}{12{,}288}
\newcommand{\sdFwdHitsFwdOcc}{94\%}
\newcommand{\sdFwdHitsFwdOccTh}{100\%}
\newcommand{\sdFwdHitsFwdIssue}{75\%}
\newcommand{\sdFwdHitsFwdSpeed}{1.26$\times$}
\newcommand{\sdOursFwdMs}{1.93}
\newcommand{\sdOursFwdSlow}{1.00$\times$}
\newcommand{\sdOursFwdDvram}{0}
\newcommand{\sdOursFwdVram}{1{,}092}
\newcommand{\sdOursFwdKeys}{1.00}
\newcommand{\sdOursFwdInstr}{682}
\newcommand{\sdOursFwdDram}{152}
\newcommand{\sdOursFwdKernels}{1}
\newcommand{\sdOursFwdRegs}{39}
\newcommand{\sdOursFwdThreads}{256}
\newcommand{\sdOursFwdSmem}{13{,}312}
\newcommand{\sdOursFwdOcc}{94\%}
\newcommand{\sdOursFwdOccTh}{100\%}
\newcommand{\sdOursFwdIssue}{82\%}
\newcommand{\sdOursFwdSpeed}{1.41$\times$}
\newcommand{\sdLossStdLossMs}{2.22}
\newcommand{\sdLossStdLossSlow}{8.05$\times$}
\newcommand{\sdLossStdLossDvram}{400}
\newcommand{\sdLossStdLossVram}{1{,}492}
\newcommand{\sdLossStdLossKeys}{1.00}
\newcommand{\sdLossStdLossInstr}{349}
\newcommand{\sdLossStdLossDram}{1{,}207}
\newcommand{\sdLossStdLossKernels}{5}
\newcommand{\sdLossStdLossRegs}{40}
\newcommand{\sdLossStdLossThreads}{256}
\newcommand{\sdLossStdLossSmem}{13{,}104}
\newcommand{\sdLossStdLossOcc}{99\%}
\newcommand{\sdLossStdLossOccTh}{100\%}
\newcommand{\sdLossStdLossIssue}{37\%}
\newcommand{\sdLossStdLossSpeed}{1.00$\times$}
\newcommand{\sdOursLossMs}{0.28}
\newcommand{\sdOursLossSlow}{1.00$\times$}
\newcommand{\sdOursLossDvram}{0}
\newcommand{\sdOursLossVram}{1{,}092}
\newcommand{\sdOursLossKeys}{1.00}
\newcommand{\sdOursLossInstr}{57}
\newcommand{\sdOursLossDram}{156}
\newcommand{\sdOursLossKernels}{1}
\newcommand{\sdOursLossRegs}{40}
\newcommand{\sdOursLossThreads}{256}
\newcommand{\sdOursLossSmem}{576}
\newcommand{\sdOursLossOcc}{97\%}
\newcommand{\sdOursLossOccTh}{100\%}
\newcommand{\sdOursLossIssue}{43\%}
\newcommand{\sdOursLossSpeed}{8.05$\times$}
\newcommand{\sdFullImagePixMs}{32.33}
\newcommand{\sdFullImagePixSlow}{3.94$\times$}
\newcommand{\sdFullImagePixDvram}{0}
\newcommand{\sdFullImagePixVram}{1{,}092}
\newcommand{\sdFullImagePixKeys}{1.00}
\newcommand{\sdFullImagePixInstr}{11{,}496}
\newcommand{\sdFullImagePixDram}{3{,}417}
\newcommand{\sdFullImagePixKernels}{12}
\newcommand{\sdFullImagePixRegs}{48}
\newcommand{\sdFullImagePixThreads}{128}
\newcommand{\sdFullImagePixSmem}{9{,}216}
\newcommand{\sdFullImagePixOcc}{83\%}
\newcommand{\sdFullImagePixOccTh}{83\%}
\newcommand{\sdFullImagePixIssue}{84\%}
\newcommand{\sdFullImagePixSpeed}{1.00$\times$}
\newcommand{\sdOursPixMs}{8.20}
\newcommand{\sdOursPixSlow}{1.00$\times$}
\newcommand{\sdOursPixDvram}{0}
\newcommand{\sdOursPixVram}{1{,}092}
\newcommand{\sdOursPixKeys}{1.00}
\newcommand{\sdOursPixInstr}{2{,}875}
\newcommand{\sdOursPixDram}{853}
\newcommand{\sdOursPixKernels}{3}
\newcommand{\sdOursPixRegs}{48}
\newcommand{\sdOursPixThreads}{128}
\newcommand{\sdOursPixSmem}{9{,}216}
\newcommand{\sdOursPixOcc}{83\%}
\newcommand{\sdOursPixOccTh}{83\%}
\newcommand{\sdOursPixIssue}{84\%}
\newcommand{\sdOursPixSpeed}{3.94$\times$}
\newcommand{\sdBwdGsplatBwdMs}{22.01}
\newcommand{\sdBwdGsplatBwdSlow}{3.68$\times$}
\newcommand{\sdBwdGsplatBwdDvram}{0}
\newcommand{\sdBwdGsplatBwdVram}{1{,}092}
\newcommand{\sdBwdGsplatBwdKeys}{1.00}
\newcommand{\sdBwdGsplatBwdInstr}{4{,}262}
\newcommand{\sdBwdGsplatBwdDram}{337}
\newcommand{\sdBwdGsplatBwdKernels}{1}
\newcommand{\sdBwdGsplatBwdRegs}{40}
\newcommand{\sdBwdGsplatBwdThreads}{256}
\newcommand{\sdBwdGsplatBwdSmem}{10{,}240}
\newcommand{\sdBwdGsplatBwdOcc}{95\%}
\newcommand{\sdBwdGsplatBwdOccTh}{100\%}
\newcommand{\sdBwdGsplatBwdIssue}{43\%}
\newcommand{\sdBwdTamingBwdMs}{19.42}
\newcommand{\sdBwdTamingBwdSlow}{3.22$\times$}
\newcommand{\sdBwdTamingBwdDvram}{620}
\newcommand{\sdBwdTamingBwdVram}{1{,}712}
\newcommand{\sdBwdTamingBwdKeys}{1.00}
\newcommand{\sdBwdTamingBwdInstr}{4{,}870}
\newcommand{\sdBwdTamingBwdDram}{1{,}039}
\newcommand{\sdBwdTamingBwdKernels}{1}
\newcommand{\sdBwdTamingBwdRegs}{58}
\newcommand{\sdBwdTamingBwdThreads}{32}
\newcommand{\sdBwdTamingBwdSmem}{772}
\newcommand{\sdBwdTamingBwdOcc}{33\%}
\newcommand{\sdBwdTamingBwdOccTh}{33\%}
\newcommand{\sdBwdTamingBwdIssue}{65\%}
\newcommand{\sdOursBwdMs}{6.00}
\newcommand{\sdOursBwdSlow}{1.00$\times$}
\newcommand{\sdOursBwdDvram}{0}
\newcommand{\sdOursBwdVram}{1{,}092}
\newcommand{\sdOursBwdKeys}{1.00}
\newcommand{\sdOursBwdInstr}{2{,}135}
\newcommand{\sdOursBwdDram}{545}
\newcommand{\sdOursBwdKernels}{1}
\newcommand{\sdOursBwdRegs}{48}
\newcommand{\sdOursBwdThreads}{128}
\newcommand{\sdOursBwdSmem}{9{,}216}
\newcommand{\sdOursBwdOcc}{83\%}
\newcommand{\sdOursBwdOccTh}{83\%}
\newcommand{\sdOursBwdIssue}{84\%}
\newcommand{\sdAdamUnfusedAdamMs}{2.07}
\newcommand{\sdAdamUnfusedAdamSlow}{3.64$\times$}
\newcommand{\sdAdamUnfusedAdamDvram}{256}
\newcommand{\sdAdamUnfusedAdamVram}{1{,}348}
\newcommand{\sdAdamUnfusedAdamKeys}{1.00}
\newcommand{\sdAdamUnfusedAdamInstr}{57}
\newcommand{\sdAdamUnfusedAdamDram}{796}
\newcommand{\sdAdamUnfusedAdamKernels}{3}
\newcommand{\sdAdamUnfusedAdamRegs}{80}
\newcommand{\sdAdamUnfusedAdamThreads}{128}
\newcommand{\sdAdamUnfusedAdamSmem}{0}
\newcommand{\sdAdamUnfusedAdamOcc}{44\%}
\newcommand{\sdAdamUnfusedAdamOccTh}{50\%}
\newcommand{\sdAdamUnfusedAdamIssue}{9\%}
\newcommand{\sdAdamDenseAdamMs}{0.91}
\newcommand{\sdAdamDenseAdamSlow}{1.58$\times$}
\newcommand{\sdAdamDenseAdamDvram}{0}
\newcommand{\sdAdamDenseAdamVram}{1{,}092}
\newcommand{\sdAdamDenseAdamKeys}{1.00}
\newcommand{\sdAdamDenseAdamInstr}{73}
\newcommand{\sdAdamDenseAdamDram}{527}
\newcommand{\sdAdamDenseAdamKernels}{1}
\newcommand{\sdAdamDenseAdamRegs}{127}
\newcommand{\sdAdamDenseAdamThreads}{128}
\newcommand{\sdAdamDenseAdamSmem}{0}
\newcommand{\sdAdamDenseAdamOcc}{28\%}
\newcommand{\sdAdamDenseAdamOccTh}{33\%}
\newcommand{\sdAdamDenseAdamIssue}{21\%}
\newcommand{\sdAdamFpThreeTwoAdamMs}{0.78}
\newcommand{\sdAdamFpThreeTwoAdamSlow}{1.36$\times$}
\newcommand{\sdAdamFpThreeTwoAdamDvram}{132}
\newcommand{\sdAdamFpThreeTwoAdamVram}{1{,}224}
\newcommand{\sdAdamFpThreeTwoAdamKeys}{1.00}
\newcommand{\sdAdamFpThreeTwoAdamInstr}{46}
\newcommand{\sdAdamFpThreeTwoAdamDram}{447}
\newcommand{\sdAdamFpThreeTwoAdamKernels}{1}
\newcommand{\sdAdamFpThreeTwoAdamRegs}{119}
\newcommand{\sdAdamFpThreeTwoAdamThreads}{128}
\newcommand{\sdAdamFpThreeTwoAdamSmem}{0}
\newcommand{\sdAdamFpThreeTwoAdamOcc}{29\%}
\newcommand{\sdAdamFpThreeTwoAdamOccTh}{33\%}
\newcommand{\sdAdamFpThreeTwoAdamIssue}{15\%}
\newcommand{\sdOursAdamMs}{0.57}
\newcommand{\sdOursAdamSlow}{1.00$\times$}
\newcommand{\sdOursAdamDvram}{0}
\newcommand{\sdOursAdamVram}{1{,}092}
\newcommand{\sdOursAdamKeys}{1.00}
\newcommand{\sdOursAdamInstr}{53}
\newcommand{\sdOursAdamDram}{319}
\newcommand{\sdOursAdamKernels}{1}
\newcommand{\sdOursAdamRegs}{127}
\newcommand{\sdOursAdamThreads}{128}
\newcommand{\sdOursAdamSmem}{0}
\newcommand{\sdOursAdamOcc}{30\%}
\newcommand{\sdOursAdamOccTh}{33\%}
\newcommand{\sdOursAdamIssue}{24\%}
\newcommand{\sdShThreeStepMs}{11.19}
\newcommand{\sdShThreeStepSlow}{1.11$\times$}
\newcommand{\sdShThreeStepDvram}{1122}
\newcommand{\sdShThreeStepVram}{2{,}214}
\newcommand{\sdShThreeStepKeys}{1.24}
\newcommand{\sdShThreeStepInstr}{3{,}156}
\newcommand{\sdShThreeStepDram}{2{,}634}
\newcommand{\sdShThreeStepKernels}{13}
\newcommand{\sdShThreeStepRegs}{48}
\newcommand{\sdShThreeStepThreads}{128}
\newcommand{\sdShThreeStepSmem}{9{,}216}
\newcommand{\sdShThreeStepOcc}{83\%}
\newcommand{\sdShThreeStepOccTh}{83\%}
\newcommand{\sdShThreeStepIssue}{84\%}
\newcommand{\sdShThreeStepSpeed}{1.00$\times$}
\newcommand{\sdOursStepMs}{10.07}
\newcommand{\sdOursStepSlow}{1.00$\times$}
\newcommand{\sdOursStepDvram}{0}
\newcommand{\sdOursStepVram}{1{,}092}
\newcommand{\sdOursStepKeys}{1.00}
\newcommand{\sdOursStepInstr}{3{,}053}
\newcommand{\sdOursStepDram}{1{,}709}
\newcommand{\sdOursStepKernels}{13}
\newcommand{\sdOursStepRegs}{48}
\newcommand{\sdOursStepThreads}{128}
\newcommand{\sdOursStepSmem}{9{,}216}
\newcommand{\sdOursStepOcc}{83\%}
\newcommand{\sdOursStepOccTh}{83\%}
\newcommand{\sdOursStepIssue}{84\%}
\newcommand{\sdOursStepSpeed}{1.11$\times$}
\newcommand{\sdPixBufMB}{150}
\newcommand{\sdPixFullExtraMB}{449}
\newcommand{\sdLadKeysTwopassShZeroSpeed}{1.00$\times$}
\newcommand{\sdLadKeysTwopassShZeroMs}{0.85}
\newcommand{\sdLadKeysTwopassShZeroDram}{339}
\newcommand{\sdLadKeysCursorNorejectShZeroSpeed}{1.27$\times$}
\newcommand{\sdLadKeysCursorNorejectShZeroMs}{0.66}
\newcommand{\sdLadKeysCursorNorejectShZeroDram}{273}
\newcommand{\sdLadKeysCursorNorejectSpeed}{1.49$\times$}
\newcommand{\sdLadKeysCursorNorejectMs}{0.57}
\newcommand{\sdLadKeysCursorNorejectDram}{221}
\newcommand{\sdLadOursSpeed}{1.74$\times$}
\newcommand{\sdLadOursMs}{0.48}
\newcommand{\sdLadOursDram}{175}
\newcommand{\sdKeysDramLessPct}{48\%\xspace}
\newcommand{\sdKeysDramLessMB}{164}
\newcommand{\sdKeysBufTwopassMB}{260}
\newcommand{\sdKeysBufOursMB}{151}
\newcommand{\sdKeysBufLessPct}{42\%\xspace}
\newcommand{\sdNumScenes}{15}

\label{sec:abl:within}

\begin{table}[t]\centering\footnotesize
\caption[Projection and keys]{Projection and keys.\ {\normalfont
From 3DGS's to ours, each row adds one change. garden on RTX~3070.}}
\setlength{\tabcolsep}{2pt}
\begin{tabular*}{\columnwidth}{@{\extracolsep{\fill}}l rrrr@{}}
\toprule
 & Kernels & DRAM Traffic & VRAM  & Time (Speedup) \\
\midrule
3DGS: two passes & 5 & 339 & 1{,}288 & 0.85 (1.00$\times$) \\
+ one fused pass & 2 & 273 & 1{,}172 & 0.66 (1.27$\times$) \\
+ direct RGB & 2 & 221 & 1{,}092 & 0.57 (1.49$\times$) \\
\textbf{+ early reject (ours)} & \textbf{2} & \textbf{175} & \textbf{1{,}092} & \textbf{0.48} (\textbf{1.74$\times$}) \\
\bottomrule
\end{tabular*}
\label{tab:sd_keys}
\end{table}

\parhead{Projection and keys.}
Fusing the two passes and tracking write offsets with a shared cursor alone is \sdLadKeysCursorNorejectShZeroSpeed{} faster than 3DGS (\cref{tab:sd_keys}). Direct RGB raises this to \sdLadKeysCursorNorejectSpeed{}, and early frustum/opacity culling using position, scale and opacity, before their
rotation and cone are loaded, to \sdLadOursSpeed{}. Together they cut DRAM traffic by \sdKeysDramLessMB{}\,MB (\sdKeysDramLessPct).
Key and sort buffers take \sdKeysBufLessPct less device memory (\sdKeysBufOursMB{} against \sdKeysBufTwopassMB{}\,MB on garden).

\parhead{Sort.}
Using 32-bit keys, the sort is \sdOursSortSpeed{} faster than with the 64-bit keys of 3DGS (\sdOursSortMs{} against \sdSortStdSortMs{}\,ms).

\begin{table}[t]\centering\footnotesize
\caption[Backward rasterization]{Backward rasterization breakdown}
\setlength{\tabcolsep}{2pt}
\begin{tabular*}{\columnwidth}{@{\extracolsep{\fill}}l rrrrrrrr@{}}
\toprule
 & Threads & Regs & Shared & Occu- & Issue & Instr. & VRAM & Time \\
 & /Block & /Thread & (KB) & pancy & Slots & (M) & (MB) & (ms) \\
\midrule
Per pixel (gsplat) & 256 & 40 & 10.0 & 95\% & 43\% & 4261 & 1{,}092 & 16.55 \\
Per splat (Taming) & 32 & 58 & 0.8 & 33\% & 65\% & 4870 & 1{,}712 & 15.32 \\
Warp per tile (LiteGS) & 128 & 56 & 8.0 & 60\% & 82\% & 1860 & 1{,}186 & 5.37 \\
\midrule
Per chunk, std.\ arith. & 128 & 62 & 11.6 & 58\% & 81\% & 2736 & 1{,}092 & 6.06 \\
\,\,+ lean arithmetic & 128 & 48 & 8.6 & 82\% & 88\% & 2486 & 1{,}092 & 5.21 \\
\,\,+ chunk culling (ours) & \textbf{128} & \textbf{48} & \textbf{9.0} & \textbf{83\%} & \textbf{84\%} & \textbf{2136} & \textbf{1{,}092} & \textbf{4.76} \\
\bottomrule
\end{tabular*}
\label{tab:bwd-ablation}
\end{table}

\parhead{Backward rasterization.}
\vet{\Cref{tab:bwd-ablation} profiles every backward kernel on the same frozen states.
A kernel's time depends on instructions executed and the GPU's instruction issue. The rate depends on occupancy: the more warps
are resident, the more often one is ready to issue while the others wait on memory. Occupancy, in turn, is set by
whichever per-SM budget a block exhausts first: its threads, its registers, or its shared memory.
gsplat's 256-thread blocks fill the warp slots (95\% occupancy), yet issue in
only 43\% of cycles, which we attribute to its atomic additions, and it executes twice our instructions.
Taming-3DGS launches one warp per block, so the limit of 16 blocks on an SM of 48 warp slots holds it at \bmTamingOcc{}
however few registers or bytes it uses.
LiteGS~\cite{liao2026litegs} executes the fewest instructions (\bmLitegsInstr{}\,M, in FP16), but 56 registers
per thread hold it at \bmLitegsOcc{}, and it is \bmLitegsSlow{} slower than ours on average.
Our per-chunk kernel with standard arithmetic is limited the same way, by \bmRegsStd{} registers per thread, uncapped, and
11.6\,KB of shared memory per block (\bmOccStd{}). The lean arithmetic moves that limit: it leaves the kernel
small enough to cap at \bmRegsLean{} registers and 8.6\,KB, so ten blocks fit on an SM (\bmOccLean{}), and it
removes \bmLeanInstr{} of the instructions as well; the \bwdAblStepLean{} comes from both. Chunk culling trades the other way: its masks
and waits lower the issue rate (88\% to 84\% of cycles) but remove \bmCullInstr{} of the instructions (\bwdAblStepSkipWait{}). With all of the above, our backward is \bwdAblSpeedOurs{} faster than gsplat's.}

\newcommand{\qaMipScenes}{7}
\newcommand{\qaTntScenes}{8}
\newcommand{\qaAllScenes}{15}
\newcommand{\qaOursMipPsnr}{28.94}
\newcommand{\qaOursMipPsnrD}{+0.00}
\newcommand{\qaOursMipPsnrSe}{0.00}
\newcommand{\qaOursMipPsnrGain}{-0.00}
\newcommand{\qaOursMipSsim}{0.855}
\newcommand{\qaOursMipSsimD}{+0.000}
\newcommand{\qaOursMipSsimSe}{0.000}
\newcommand{\qaOursMipLpips}{0.303}
\newcommand{\qaOursMipLpipsD}{+0.000}
\newcommand{\qaOursMipLpipsSe}{0.000}
\newcommand{\qaOursMipSplats}{1.55}
\newcommand{\qaOursMipMb}{112}
\newcommand{\qaOursMipMin}{4.4}
\newcommand{\qaOursMipRuns}{98}
\newcommand{\qaFwdStdMipPsnr}{28.96}
\newcommand{\qaFwdStdMipPsnrD}{+0.02}
\newcommand{\qaFwdStdMipPsnrSe}{0.01}
\newcommand{\qaFwdStdMipPsnrGain}{-0.02}
\newcommand{\qaFwdStdMipSsim}{0.854}
\newcommand{\qaFwdStdMipSsimD}{$-$0.000}
\newcommand{\qaFwdStdMipSsimSe}{0.000}
\newcommand{\qaFwdStdMipLpips}{0.303}
\newcommand{\qaFwdStdMipLpipsD}{+0.000}
\newcommand{\qaFwdStdMipLpipsSe}{0.000}
\newcommand{\qaFwdStdMipSplats}{1.52}
\newcommand{\qaFwdStdMipMb}{110}
\newcommand{\qaFwdStdMipMin}{14.8}
\newcommand{\qaFwdStdMipRuns}{97}
\newcommand{\qaLossStdMipPsnr}{28.79}
\newcommand{\qaLossStdMipPsnrD}{$-$0.15}
\newcommand{\qaLossStdMipPsnrSe}{0.05}
\newcommand{\qaLossStdMipPsnrGain}{0.15}
\newcommand{\qaLossStdMipSsim}{0.854}
\newcommand{\qaLossStdMipSsimD}{$-$0.001}
\newcommand{\qaLossStdMipSsimSe}{0.002}
\newcommand{\qaLossStdMipLpips}{0.312}
\newcommand{\qaLossStdMipLpipsD}{+0.009}
\newcommand{\qaLossStdMipLpipsSe}{0.006}
\newcommand{\qaLossStdMipSplats}{1.35}
\newcommand{\qaLossStdMipMb}{97}
\newcommand{\qaLossStdMipMin}{5.1}
\newcommand{\qaLossStdMipRuns}{97}
\newcommand{\qaShthreeMipPsnr}{28.68}
\newcommand{\qaShthreeMipPsnrD}{$-$0.26}
\newcommand{\qaShthreeMipPsnrSe}{0.16}
\newcommand{\qaShthreeMipPsnrGain}{0.26}
\newcommand{\qaShthreeMipSsim}{0.855}
\newcommand{\qaShthreeMipSsimD}{+0.000}
\newcommand{\qaShthreeMipSsimSe}{0.002}
\newcommand{\qaShthreeMipLpips}{0.304}
\newcommand{\qaShthreeMipLpipsD}{+0.001}
\newcommand{\qaShthreeMipLpipsSe}{0.001}
\newcommand{\qaShthreeMipSplats}{1.29}
\newcommand{\qaShthreeMipMb}{305}
\newcommand{\qaShthreeMipMin}{4.5}
\newcommand{\qaShthreeMipRuns}{97}
\newcommand{\qaKeysStdMipPsnr}{28.94}
\newcommand{\qaKeysStdMipPsnrD}{+0.00}
\newcommand{\qaKeysStdMipPsnrSe}{0.02}
\newcommand{\qaKeysStdMipPsnrGain}{-0.00}
\newcommand{\qaKeysStdMipSsim}{0.855}
\newcommand{\qaKeysStdMipSsimD}{+0.000}
\newcommand{\qaKeysStdMipSsimSe}{0.000}
\newcommand{\qaKeysStdMipLpips}{0.303}
\newcommand{\qaKeysStdMipLpipsD}{$-$0.000}
\newcommand{\qaKeysStdMipLpipsSe}{0.000}
\newcommand{\qaKeysStdMipSplats}{1.56}
\newcommand{\qaKeysStdMipRuns}{97}
\newcommand{\qaBwdGsplatMipPsnr}{28.95}
\newcommand{\qaBwdGsplatMipPsnrD}{+0.01}
\newcommand{\qaBwdGsplatMipPsnrSe}{0.01}
\newcommand{\qaBwdGsplatMipPsnrGain}{-0.01}
\newcommand{\qaBwdGsplatMipSsim}{0.855}
\newcommand{\qaBwdGsplatMipSsimD}{$-$0.000}
\newcommand{\qaBwdGsplatMipSsimSe}{0.000}
\newcommand{\qaBwdGsplatMipLpips}{0.303}
\newcommand{\qaBwdGsplatMipLpipsD}{+0.000}
\newcommand{\qaBwdGsplatMipLpipsSe}{0.000}
\newcommand{\qaBwdGsplatMipSplats}{1.56}
\newcommand{\qaBwdGsplatMipRuns}{97}
\newcommand{\qaAdamStdMipPsnr}{28.95}
\newcommand{\qaAdamStdMipPsnrD}{+0.01}
\newcommand{\qaAdamStdMipPsnrSe}{0.01}
\newcommand{\qaAdamStdMipPsnrGain}{-0.01}
\newcommand{\qaAdamStdMipSsim}{0.854}
\newcommand{\qaAdamStdMipSsimD}{$-$0.000}
\newcommand{\qaAdamStdMipSsimSe}{0.000}
\newcommand{\qaAdamStdMipLpips}{0.303}
\newcommand{\qaAdamStdMipLpipsD}{+0.000}
\newcommand{\qaAdamStdMipLpipsSe}{0.000}
\newcommand{\qaAdamStdMipSplats}{1.55}
\newcommand{\qaAdamStdMipRuns}{97}
\newcommand{\qaOursTntPsnr}{24.33}
\newcommand{\qaOursTntPsnrD}{+0.00}
\newcommand{\qaOursTntPsnrSe}{0.00}
\newcommand{\qaOursTntPsnrGain}{-0.00}
\newcommand{\qaOursTntSsim}{0.813}
\newcommand{\qaOursTntSsimD}{+0.000}
\newcommand{\qaOursTntSsimSe}{0.000}
\newcommand{\qaOursTntLpips}{0.272}
\newcommand{\qaOursTntLpipsD}{+0.000}
\newcommand{\qaOursTntLpipsSe}{0.000}
\newcommand{\qaOursTntSplats}{1.72}
\newcommand{\qaOursTntMb}{124}
\newcommand{\qaOursTntMin}{4.6}
\newcommand{\qaOursTntRuns}{113}
\newcommand{\qaFwdStdTntPsnr}{24.35}
\newcommand{\qaFwdStdTntPsnrD}{+0.01}
\newcommand{\qaFwdStdTntPsnrSe}{0.02}
\newcommand{\qaFwdStdTntPsnrGain}{-0.01}
\newcommand{\qaFwdStdTntSsim}{0.813}
\newcommand{\qaFwdStdTntSsimD}{+0.000}
\newcommand{\qaFwdStdTntSsimSe}{0.000}
\newcommand{\qaFwdStdTntLpips}{0.272}
\newcommand{\qaFwdStdTntLpipsD}{$-$0.000}
\newcommand{\qaFwdStdTntLpipsSe}{0.000}
\newcommand{\qaFwdStdTntSplats}{1.72}
\newcommand{\qaFwdStdTntMb}{124}
\newcommand{\qaFwdStdTntMin}{15.3}
\newcommand{\qaFwdStdTntRuns}{113}
\newcommand{\qaLossStdTntPsnr}{24.20}
\newcommand{\qaLossStdTntPsnrD}{$-$0.13}
\newcommand{\qaLossStdTntPsnrSe}{0.06}
\newcommand{\qaLossStdTntPsnrGain}{0.13}
\newcommand{\qaLossStdTntSsim}{0.815}
\newcommand{\qaLossStdTntSsimD}{+0.003}
\newcommand{\qaLossStdTntSsimSe}{0.001}
\newcommand{\qaLossStdTntLpips}{0.271}
\newcommand{\qaLossStdTntLpipsD}{$-$0.001}
\newcommand{\qaLossStdTntLpipsSe}{0.001}
\newcommand{\qaLossStdTntSplats}{1.75}
\newcommand{\qaLossStdTntMb}{126}
\newcommand{\qaLossStdTntMin}{5.3}
\newcommand{\qaLossStdTntRuns}{113}
\newcommand{\qaShthreeTntPsnr}{23.62}
\newcommand{\qaShthreeTntPsnrD}{$-$0.71}
\newcommand{\qaShthreeTntPsnrSe}{0.11}
\newcommand{\qaShthreeTntPsnrGain}{0.71}
\newcommand{\qaShthreeTntSsim}{0.806}
\newcommand{\qaShthreeTntSsimD}{$-$0.007}
\newcommand{\qaShthreeTntSsimSe}{0.001}
\newcommand{\qaShthreeTntLpips}{0.284}
\newcommand{\qaShthreeTntLpipsD}{+0.011}
\newcommand{\qaShthreeTntLpipsSe}{0.002}
\newcommand{\qaShthreeTntSplats}{1.08}
\newcommand{\qaShthreeTntMb}{254}
\newcommand{\qaShthreeTntMin}{4.3}
\newcommand{\qaShthreeTntRuns}{113}
\newcommand{\qaKeysStdTntPsnr}{24.32}
\newcommand{\qaKeysStdTntPsnrD}{$-$0.01}
\newcommand{\qaKeysStdTntPsnrSe}{0.01}
\newcommand{\qaKeysStdTntPsnrGain}{0.01}
\newcommand{\qaKeysStdTntSsim}{0.812}
\newcommand{\qaKeysStdTntSsimD}{$-$0.000}
\newcommand{\qaKeysStdTntSsimSe}{0.000}
\newcommand{\qaKeysStdTntLpips}{0.272}
\newcommand{\qaKeysStdTntLpipsD}{+0.000}
\newcommand{\qaKeysStdTntLpipsSe}{0.000}
\newcommand{\qaKeysStdTntSplats}{1.72}
\newcommand{\qaKeysStdTntRuns}{113}
\newcommand{\qaBwdGsplatTntPsnr}{24.34}
\newcommand{\qaBwdGsplatTntPsnrD}{+0.00}
\newcommand{\qaBwdGsplatTntPsnrSe}{0.01}
\newcommand{\qaBwdGsplatTntPsnrGain}{-0.00}
\newcommand{\qaBwdGsplatTntSsim}{0.813}
\newcommand{\qaBwdGsplatTntSsimD}{+0.000}
\newcommand{\qaBwdGsplatTntSsimSe}{0.000}
\newcommand{\qaBwdGsplatTntLpips}{0.272}
\newcommand{\qaBwdGsplatTntLpipsD}{$-$0.000}
\newcommand{\qaBwdGsplatTntLpipsSe}{0.000}
\newcommand{\qaBwdGsplatTntSplats}{1.72}
\newcommand{\qaBwdGsplatTntRuns}{113}
\newcommand{\qaAdamStdTntPsnr}{24.34}
\newcommand{\qaAdamStdTntPsnrD}{+0.01}
\newcommand{\qaAdamStdTntPsnrSe}{0.02}
\newcommand{\qaAdamStdTntPsnrGain}{-0.01}
\newcommand{\qaAdamStdTntSsim}{0.813}
\newcommand{\qaAdamStdTntSsimD}{$-$0.000}
\newcommand{\qaAdamStdTntSsimSe}{0.000}
\newcommand{\qaAdamStdTntLpips}{0.272}
\newcommand{\qaAdamStdTntLpipsD}{$-$0.000}
\newcommand{\qaAdamStdTntLpipsSe}{0.000}
\newcommand{\qaAdamStdTntSplats}{1.72}
\newcommand{\qaAdamStdTntRuns}{113}
\newcommand{\qaOursAllPsnr}{26.48}
\newcommand{\qaOursAllPsnrD}{+0.00}
\newcommand{\qaOursAllPsnrSe}{0.00}
\newcommand{\qaOursAllPsnrGain}{-0.00}
\newcommand{\qaOursAllSsim}{0.832}
\newcommand{\qaOursAllSsimD}{+0.000}
\newcommand{\qaOursAllSsimSe}{0.000}
\newcommand{\qaOursAllLpips}{0.287}
\newcommand{\qaOursAllLpipsD}{+0.000}
\newcommand{\qaOursAllLpipsSe}{0.000}
\newcommand{\qaOursAllSplats}{1.64}
\newcommand{\qaOursAllMb}{118}
\newcommand{\qaOursAllMin}{4.5}
\newcommand{\qaOursAllRuns}{211}
\newcommand{\qaFwdStdAllPsnr}{26.50}
\newcommand{\qaFwdStdAllPsnrD}{+0.02}
\newcommand{\qaFwdStdAllPsnrSe}{0.01}
\newcommand{\qaFwdStdAllPsnrGain}{-0.02}
\newcommand{\qaFwdStdAllSsim}{0.832}
\newcommand{\qaFwdStdAllSsimD}{+0.000}
\newcommand{\qaFwdStdAllSsimSe}{0.000}
\newcommand{\qaFwdStdAllLpips}{0.286}
\newcommand{\qaFwdStdAllLpipsD}{$-$0.000}
\newcommand{\qaFwdStdAllLpipsSe}{0.000}
\newcommand{\qaFwdStdAllSplats}{1.63}
\newcommand{\qaFwdStdAllMb}{117}
\newcommand{\qaFwdStdAllMin}{15.1}
\newcommand{\qaFwdStdAllRuns}{210}
\newcommand{\qaLossStdAllPsnr}{26.34}
\newcommand{\qaLossStdAllPsnrD}{$-$0.14}
\newcommand{\qaLossStdAllPsnrSe}{0.04}
\newcommand{\qaLossStdAllPsnrGain}{0.14}
\newcommand{\qaLossStdAllSsim}{0.833}
\newcommand{\qaLossStdAllSsimD}{+0.001}
\newcommand{\qaLossStdAllSsimSe}{0.001}
\newcommand{\qaLossStdAllLpips}{0.290}
\newcommand{\qaLossStdAllLpipsD}{+0.003}
\newcommand{\qaLossStdAllLpipsSe}{0.003}
\newcommand{\qaLossStdAllSplats}{1.56}
\newcommand{\qaLossStdAllMb}{112}
\newcommand{\qaLossStdAllMin}{5.2}
\newcommand{\qaLossStdAllRuns}{210}
\newcommand{\qaShthreeAllPsnr}{25.99}
\newcommand{\qaShthreeAllPsnrD}{$-$0.50}
\newcommand{\qaShthreeAllPsnrSe}{0.11}
\newcommand{\qaShthreeAllPsnrGain}{0.50}
\newcommand{\qaShthreeAllSsim}{0.829}
\newcommand{\qaShthreeAllSsimD}{$-$0.003}
\newcommand{\qaShthreeAllSsimSe}{0.001}
\newcommand{\qaShthreeAllLpips}{0.293}
\newcommand{\qaShthreeAllLpipsD}{+0.007}
\newcommand{\qaShthreeAllLpipsSe}{0.002}
\newcommand{\qaShthreeAllSplats}{1.18}
\newcommand{\qaShthreeAllMb}{278}
\newcommand{\qaShthreeAllMin}{4.4}
\newcommand{\qaShthreeAllRuns}{210}
\newcommand{\qaKeysStdAllPsnr}{26.48}
\newcommand{\qaKeysStdAllPsnrD}{$-$0.01}
\newcommand{\qaKeysStdAllPsnrSe}{0.01}
\newcommand{\qaKeysStdAllPsnrGain}{0.01}
\newcommand{\qaKeysStdAllSsim}{0.832}
\newcommand{\qaKeysStdAllSsimD}{$-$0.000}
\newcommand{\qaKeysStdAllSsimSe}{0.000}
\newcommand{\qaKeysStdAllLpips}{0.287}
\newcommand{\qaKeysStdAllLpipsD}{+0.000}
\newcommand{\qaKeysStdAllLpipsSe}{0.000}
\newcommand{\qaKeysStdAllSplats}{1.64}
\newcommand{\qaKeysStdAllRuns}{210}
\newcommand{\qaBwdGsplatAllPsnr}{26.49}
\newcommand{\qaBwdGsplatAllPsnrD}{+0.01}
\newcommand{\qaBwdGsplatAllPsnrSe}{0.01}
\newcommand{\qaBwdGsplatAllPsnrGain}{-0.01}
\newcommand{\qaBwdGsplatAllSsim}{0.832}
\newcommand{\qaBwdGsplatAllSsimD}{+0.000}
\newcommand{\qaBwdGsplatAllSsimSe}{0.000}
\newcommand{\qaBwdGsplatAllLpips}{0.287}
\newcommand{\qaBwdGsplatAllLpipsD}{$-$0.000}
\newcommand{\qaBwdGsplatAllLpipsSe}{0.000}
\newcommand{\qaBwdGsplatAllSplats}{1.64}
\newcommand{\qaBwdGsplatAllRuns}{210}
\newcommand{\qaAdamStdAllPsnr}{26.49}
\newcommand{\qaAdamStdAllPsnrD}{+0.01}
\newcommand{\qaAdamStdAllPsnrSe}{0.01}
\newcommand{\qaAdamStdAllPsnrGain}{-0.01}
\newcommand{\qaAdamStdAllSsim}{0.832}
\newcommand{\qaAdamStdAllSsimD}{$-$0.000}
\newcommand{\qaAdamStdAllSsimSe}{0.000}
\newcommand{\qaAdamStdAllLpips}{0.287}
\newcommand{\qaAdamStdAllLpipsD}{+0.000}
\newcommand{\qaAdamStdAllLpipsSe}{0.000}
\newcommand{\qaAdamStdAllSplats}{1.64}
\newcommand{\qaAdamStdAllRuns}{210}
\newcommand{\qaCheckMaxPsnrD}{0.01}
\newcommand{\qaRunsMin}{97}
\newcommand{\qaRunsPerSceneMin}{12}

\newcommand{\qcLocalScale}{0.86}

\subsection{Performance--Quality Trade-offs}
\label{sec:abl:quality}
We investigate the performance--quality trade-offs (\cref{tab:stage_times}) of the three changes that could affect quality: strided pixel sampling, flat RGB color with directional opacity, and Tile SSIM.

\papfig{Quality against training time}{diagrams/fig_quality_curve}
\parhead{Strided pixel sampling.}
Strided pixel sampling trains on a quarter of the pixels per step. Yet supervising every pixel instead reaches the same quality (\stTabFwdStdDPsnr\,dB PSNR, \stTabFwdStdDSsim{} SSIM) while training \stFwdStdStepX$\times$ longer (together with the standard forward, \cref{tab:stage_times}). \Cref{fig:quality_curve} (left) also contrasts strided sampling with training at half resolution, which touches the same number of pixels per step: training time is similar, but half resolution loses quality.


\parhead{Directional opacity and RGB\@.}
Reverting RGB and cone opacity to SH degree~3 without directional opacity shows that our change slows training slightly (\stSHThreeStepX$\times$), mainly because it keeps more splats. At an equal splat count, the bandwidth-bound stages should make ours faster. \Cref{tab:stage_times} shows this too: the projection forward and projection backward + Adam stages get faster, while the other stages slow down slightly with the higher splat count. Overall, our model improves PSNR by \qaShthreeAllPsnrGain\,dB over SH degree~3, and SSIM and LPIPS improve as well. The split between base and cone splats is given in \cref{sec:results:main}.

\papfig{Qualitative comparison on held-out views}{diagrams/fig_qualitative}
\Cref{fig:qualitative} shows selected views where our model captures reflections and specularities much better, and with better contrast. With directional opacity, splats are no longer tied to geometry, since they can vanish completely at view angles where they do not belong. Instead, they appear to model light and lighting effects.

\papfig{Failure cases of directional opacity}{diagrams/fig_cone_failure}
\textit{Failure cases.} At view angles far from the training cameras, cone splats sometimes produce extreme artifacts (\cref{fig:cone_failure}). The test views do not show them: the test set is every 8th view, so each test view stays close to its neighbouring training views.

\parhead{Tile SSIM\@.}
Replacing the full \texttt{fused-ssim} with Tile SSIM makes training \stLossStdStepX$\times$ faster (\cref{tab:stage_times}) with almost no change in quality. The gain comes entirely from the loss stage, which becomes \stLossStdLossX$\times$ faster. \Cref{fig:quality_curve} (right) compares Tile SSIM with \texttt{fused-ssim} (w/o Tile SSIM) on the standard SSIM metric. Even on that metric, ours follows closely, and arguably does better when plotted against time rather than iterations.

\section{Conclusion and Discussion}
\label{sec:conclusion}

We presented \method{}, a system that trains 3DGS at full resolution on consumer GPUs by removing redundant work from every stage of a training step. It supervises a strided quarter of the full-resolution pixels per step with a single-pass tile SSIM loss, represents each splat with a flat RGB color and a directional cone opacity, fuses projection with key emission and backward projection with a sparse Adam step over compact moments, and gives the backward rasterization one warp per splat.  At the \num{30}k-step budget, it is the fastest of \cmpMethods{} methods on both datasets and all four GPUs, at least \cmpGainShortMin$\times$ faster than the quickest of the others, with competitive quality. On an \cmpEntryCard{}, it trains a full-resolution Mip-NeRF~360 scene in \cmpOursShortEntryTime{}, faster than any other method trains it on an RTX~4090.

\method{} has three main limitations. First, cone splats can produce strong artifacts at view angles far from the training cameras (\cref{fig:cone_failure}); the standard test split, every eighth view, does not show them. We plan to regularize the cones to avoid them. Second, strided pixel sampling relies on neighboring pixels being similar, which holds at full resolution but less so at lower resolutions. Third, we evaluated only NVIDIA GPUs; supporting other consumer hardware, such as AMD GPUs and Apple silicon, is future work.



\bibliographystyle{ACM-Reference-Format}
\bibliography{main}

@article{kerbl20233dgs,
  author    = {Kerbl, Bernhard and Kopanas, Georgios and Leimk{\"u}hler, Thomas and Drettakis, George},
  title     = {{3D} Gaussian Splatting for Real-Time Radiance Field Rendering},
  journal   = {ACM Transactions on Graphics},
  volume    = {42},
  number    = {4},
  year      = {2023},
  articleno = {139},
  numpages  = {14},
}

@inproceedings{mallick2024taming,
  author    = {Mallick, Saswat Subhajyoti and Goel, Rahul and Kerbl, Bernhard and Steinberger, Markus and Carrasco, Francisco Vicente and De La Torre, Fernando},
  title     = {Taming {3DGS}: High-Quality Radiance Fields with Limited Resources},
  booktitle = {SIGGRAPH Asia 2024 Conference Papers},
  year      = {2024},
  publisher = {ACM},
  address   = {Tokyo, Japan},
  articleno = {2},
  numpages  = {11},
}

@article{ye2024gsplat,
  author  = {Ye, Vickie and Li, Ruilong and Kerr, Justin and Turkulainen, Matias and Yi, Brent and Pan, Zhuoyang and Seiskari, Otto and Ye, Jianbo and Hu, Jeffrey and Tancik, Matthew and Kanazawa, Angjoo},
  title   = {gsplat: An Open-Source Library for {G}aussian Splatting},
  journal = {Journal of Machine Learning Research},
  volume  = {26},
  number  = {34},
  pages   = {1--17},
  year    = {2025},
}

@inproceedings{barron2022mipnerf360,
  author    = {Barron, Jonathan T. and Mildenhall, Ben and Verbin, Dor and Srinivasan, Pratul P. and Hedman, Peter},
  title     = {Mip-{NeRF} 360: Unbounded Anti-Aliased Neural Radiance Fields},
  booktitle = {CVPR},
  year      = {2022},
  publisher = {IEEE},
  address   = {New Orleans, LA, USA},
  pages     = {5460--5469},
}

@article{knapitsch2017tanks,
  author    = {Knapitsch, Arno and Park, Jaesik and Zhou, Qian-Yi and Koltun, Vladlen},
  title     = {Tanks and Temples: Benchmarking Large-Scale Scene Reconstruction},
  journal   = {ACM Transactions on Graphics},
  volume    = {36},
  number    = {4},
  year      = {2017},
  articleno = {78},
  numpages  = {13},
}

@article{wang2004image,
  author  = {Wang, Zhou and Bovik, Alan C. and Sheikh, Hamid R. and Simoncelli, Eero P.},
  title   = {Image Quality Assessment: From Error Visibility to Structural Similarity},
  journal = {IEEE Transactions on Image Processing},
  volume  = {13},
  number  = {4},
  year    = {2004},
  pages   = {600--612},
}

@inproceedings{zhang2018unreasonable,
  author    = {Zhang, Richard and Isola, Phillip and Efros, Alexei A. and Shechtman, Eli and Wang, Oliver},
  title     = {The Unreasonable Effectiveness of Deep Features as a Perceptual Metric},
  booktitle = {CVPR},
  year      = {2018},
  publisher = {IEEE},
  address   = {Salt Lake City, UT, USA},
  pages     = {586--595},
}

@misc{durvasula2023distwar,
  author       = {Durvasula, Sankeerth and Zhao, Adrian and Chen, Fan and Liang, Ruofan and Sanjaya, Pawan Kumar and Vijaykumar, Nandita},
  title        = {{DISTWAR}: Fast Differentiable Rendering on Raster-based Rendering Pipelines},
  year         = {2023},
  howpublished = {arXiv:2401.05345},
}

@article{radl2024stopthepop,
  author    = {Radl, Lukas and Steiner, Michael and Parger, Mathias and Weinrauch, Alexander and Kerbl, Bernhard and Steinberger, Markus},
  title     = {{StopThePop}: Sorted {G}aussian Splatting for View-Consistent Real-time Rendering},
  year      = {2024},
  journal   = {ACM Transactions on Graphics},
  volume    = {43},
  number    = {4},
  articleno = {64},
  numpages  = {17},
}

@inproceedings{kheradmand20243dgs,
  author    = {Kheradmand, Shakiba and Rebain, Daniel and Sharma, Gopal and Sun, Weiwei and Tseng, Yang-Che and Isack, Hossam and Kar, Abhishek and Tagliasacchi, Andrea and Yi, Kwang Moo},
  title     = {{3D} {G}aussian Splatting as {M}arkov Chain {M}onte {C}arlo},
  booktitle = {NeurIPS},
  year      = {2024},
  publisher = {Curran Associates, Inc.},
  address   = {Vancouver, BC, Canada},
  pages     = {80965--80986},
}

@inproceedings{hahlbohm2026fastergs,
  author    = {Hahlbohm, Florian and Franke, Linus and Eisemann, Martin and Magnor, Marcus},
  title     = {{Faster-GS}: Analyzing and Improving {Gaussian} Splatting Optimization},
  booktitle = {CVPR},
  year      = {2026},
  note      = {arXiv:2602.09999},
  publisher = {IEEE/CVF},
  address   = {Denver, CO, USA},
  pages     = {18946--18957},
}

@inproceedings{fastgs,
  author    = {Ren, Shiwei and Wen, Tianci and Fang, Yongchun and Lu, Biao},
  title     = {{FastGS}: Training {3D} Gaussian Splatting in 100 Seconds},
  booktitle = {CVPR},
  year      = {2026},
  note      = {arXiv:2511.04283},
  publisher = {IEEE/CVF},
  address   = {Denver, CO, USA},
  pages     = {26094--26103},
}

@inproceedings{hanson2025speedysplat,
  author    = {Hanson, Alex and Tu, Allen and Lin, Geng and Singla, Vasu and Zwicker, Matthias and Goldstein, Tom},
  title     = {Speedy-Splat: Fast {3D} Gaussian Splatting with Sparse Pixels and Sparse Primitives},
  booktitle = {CVPR},
  year      = {2025},
  note      = {arXiv:2412.00578},
  publisher = {IEEE},
  address   = {Nashville, TN, USA},
  pages     = {21537--21546},
}

@inproceedings{chen2025dashgaussian,
  author    = {Chen, Youyu and Jiang, Junjun and Jiang, Kui and Tang, Xiao and Li, Zhihao and Liu, Xianming and Nie, Yinyu},
  title     = {DashGaussian: Optimizing {3D} Gaussian Splatting in 200 Seconds},
  booktitle = {CVPR},
  year      = {2025},
  note      = {arXiv:2503.18402},
  publisher = {IEEE},
  address   = {Nashville, TN, USA},
  pages     = {11146--11155},
}

@inproceedings{dhiman2024turbogs,
  author    = {Dhiman, Ankit and Lu, Tao and Srinath, R and Arslan, Emre and Xing, Angela and Xiangli, Yuanbo and Radhakrishnan, Venkatesh Babu and Sridhar, Srinath},
  title     = {Turbo-{GS}: Accelerating {3D} Gaussian Fitting for High-Resolution Radiance Fields},
  booktitle = {CVPR},
  year      = {2026},
  publisher = {IEEE/CVF},
  address   = {Denver, CO, USA},
  pages     = {15454--15464},
}

@article{fang2024minisplatting2,
  author  = {Fang, Guangchi and Wang, Bing},
  title   = {Efficient Scene Modeling via Structure-Aware and Region-Prioritized {3D} Gaussians},
  journal = {IEEE Transactions on Pattern Analysis and Machine Intelligence},
  year    = {2026},
  volume  = {48},
  number  = {4},
  pages   = {4623--4641},
}

@inproceedings{hollein2024gslm,
  author    = {H{\"o}llein, Lukas and Bo{\v z}i{\v c}, Alja{\v z} and Zollh{\"o}fer, Michael and Nie{\ss}ner, Matthias},
  title     = {{3DGS-LM}: Faster Gaussian-Splatting Optimization with Levenberg-Marquardt},
  booktitle = {ICCV},
  year      = {2025},
  note      = {arXiv:2409.12892},
  publisher = {IEEE},
  address   = {Honolulu, HI, USA},
  pages     = {26740--26750},
}

@inproceedings{zhao2025grendel,
  author    = {Zhao, Hexu and Weng, Haoyang and Lu, Daohan and Li, Ang and Li, Jinyang and Panda, Aurojit and Xie, Saining},
  title     = {On Scaling Up {3D} Gaussian Splatting Training},
  booktitle = {ICLR},
  year      = {2025},
  note      = {arXiv:2406.18533},
  publisher = {OpenReview.net},
  address   = {Singapore},
  numpages  = {20},
}

@inproceedings{yu2024mipsplatting,
  author    = {Yu, Zehao and Chen, Anpei and Huang, Binbin and Sattler, Torsten and Geiger, Andreas},
  title     = {Mip-Splatting: Alias-free {3D} Gaussian Splatting},
  booktitle = {CVPR},
  year      = {2024},
  publisher = {IEEE},
  address   = {Seattle, WA, USA},
  pages     = {19447--19456},
}

@inproceedings{yan2023multiscale,
  author    = {Yan, Zhiwen and Low, Weng Fei and Chen, Yu and Lee, Gim Hee},
  title     = {Multi-Scale {3D} Gaussian Splatting for Anti-Aliased Rendering},
  booktitle = {CVPR},
  year      = {2024},
  publisher = {IEEE},
  address   = {Seattle, WA, USA},
  pages     = {20923--20931},
}

@article{kerbl2024hierarchical,
  author    = {Kerbl, Bernhard and Meuleman, Andreas and Kopanas, Georgios and Wimmer, Michael and Lanvin, Alexandre and Drettakis, George},
  title     = {A Hierarchical {3D} Gaussian Representation for Real-Time Rendering of Very Large Datasets},
  journal   = {ACM Transactions on Graphics},
  volume    = {43},
  number    = {4},
  year      = {2024},
  articleno = {62},
  numpages  = {15},
}

@inproceedings{lin2024vastgaussian,
  author    = {Lin, Jiaqi and Li, Zhihao and Tang, Xiao and Liu, Jianzhuang and Liu, Shiyong and Liu, Jiayue and Lu, Yangdi and Wu, Xiaofei and Xu, Songcen and Yan, Youliang and Yang, Wenming},
  title     = {VastGaussian: Vast {3D} Gaussians for Large Scene Reconstruction},
  booktitle = {CVPR},
  year      = {2024},
  publisher = {IEEE},
  address   = {Seattle, WA, USA},
  pages     = {5166--5175},
}

@inproceedings{liu2024citygaussian,
  author    = {Liu, Yang and Guan, He and Luo, Chuanchen and Fan, Lue and Wang, Naiyan and Peng, Junran and Zhang, Zhaoxiang},
  title     = {CityGaussian: Real-time High-quality Large-Scale Scene Rendering with Gaussians},
  booktitle = {ECCV},
  year      = {2024},
  publisher = {Springer},
  address   = {Milan, Italy},
  pages     = {265--282},
}

@misc{li2025hrgs,
  author       = {Li, Changbai and Zhu, Haodong and Chen, Hanlin and Zhang, Juan and Chen, Tongfei and Yang, Shuo and Shao, Shuwei and Dong, Wenhao and Zhang, Baochang},
  title        = {{HRGS}: Hierarchical Gaussian Splatting for Memory-Efficient High-Resolution {3D} Reconstruction},
  year         = {2025},
  howpublished = {arXiv:2506.14229},
}

@inproceedings{girish2024eagles,
  author    = {Girish, Sharath and Gupta, Kamal and Shrivastava, Abhinav},
  title     = {{EAGLES}: Efficient Accelerated {3D} Gaussians with Lightweight EncodingS},
  booktitle = {ECCV},
  year      = {2024},
  publisher = {Springer},
  address   = {Milan, Italy},
  pages     = {54--71},
}

@inproceedings{lee2024compact,
  author    = {Lee, Joo Chan and Rho, Daniel and Sun, Xiangyu and Ko, Jong Hwan and Park, Eunbyung},
  title     = {Compact {3D} Gaussian Representation for Radiance Field},
  booktitle = {CVPR},
  year      = {2024},
  publisher = {IEEE},
  address   = {Seattle, WA, USA},
  pages     = {21719--21728},
}

@article{papantonakis2024reducing,
  author    = {Papantonakis, Panagiotis and Kopanas, Georgios and Kerbl, Bernhard and Lanvin, Alexandre and Drettakis, George},
  title     = {Reducing the Memory Footprint of {3D} Gaussian Splatting},
  journal   = {Proceedings of the ACM on Computer Graphics and Interactive Techniques},
  volume    = {7},
  number    = {1},
  year      = {2024},
  articleno = {16},
  numpages  = {17},
}

@inproceedings{lee2025gsscale,
  author    = {Lee, Donghyun and Jeong, Dawoon and Lee, Jae W. and Yoon, Hongil},
  title     = {{GS-Scale}: Unlocking Large-Scale {3D} Gaussian Splatting Training via Host Offloading},
  year      = {2026},
  booktitle = {ASPLOS},
  publisher = {ACM},
  address   = {Pittsburgh, PA, USA},
  pages     = {860--875},
}

@inproceedings{zhao2026clm,
  author    = {Zhao, Hexu and Min, Xiwen and Liu, Xiaoteng and Gong, Moonjun and Li, Yiming and Li, Ang and Xie, Saining and Li, Jinyang and Panda, Aurojit},
  title     = {{CLM}: Removing the {GPU} Memory Barrier for {3D} Gaussian Splatting},
  booktitle = {ASPLOS},
  year      = {2026},
  note      = {arXiv:2511.04951},
  publisher = {ACM},
  address   = {Pittsburgh, PA, USA},
  pages     = {377--393},
}

@inproceedings{vksplat,
  booktitle = {Eurographics 2026 - Short Papers},
  title     = {{VkSplat: High-Performance 3DGS Training in Vulkan Compute}},
  author    = {Chen, Jingxiang and Ibrahim, Mohamed and Liu, Yang},
  year      = {2026},
  publisher = {The Eurographics Association},
  issn      = {2309-5059},
  isbn      = {978-3-03868-299-8},
  doi       = {10.2312/egs.20261024},
  address   = {Aachen, Germany},
  numpages  = {4},
}

@inproceedings{zhang2024pixelgs,
  author    = {Zhang, Zheng and Hu, Wenbo and Lao, Yixing and He, Tong and Zhao, Hengshuang},
  title     = {Pixel-{GS}: Density Control with Pixel-Aware Gradient for {3D} Gaussian Splatting},
  booktitle = {ECCV},
  year      = {2024},
  publisher = {Springer},
  address   = {Milan, Italy},
  pages     = {326--342},
}

@inproceedings{bulo2024revising,
  author    = {Rota Bul{\`o}, Samuel and Porzi, Lorenzo and Kontschieder, Peter},
  title     = {Revising Densification in Gaussian Splatting},
  booktitle = {ECCV},
  year      = {2024},
  publisher = {Springer},
  address   = {Milan, Italy},
  pages     = {347--362},
}

@inproceedings{grubert2025improving,
  author    = {Grubert, Glenn and Barthel, Florian and Hilsmann, Anna and Eisert, Peter},
  title     = {Improving Adaptive Density Control for {3D} Gaussian Splatting},
  booktitle = {VISIGRAPP},
  year      = {2025},
  publisher = {SciTePress},
  address   = {Porto, Portugal},
  pages     = {610--621},
}

@misc{elrawy2025opacity,
  author       = {Elrawy, Abdelrhman and Mohammed, Emad A.},
  title        = {Opacity-Gradient Driven Density Control for Compact and Efficient Few-Shot {3D} Gaussian Splatting},
  year         = {2025},
  howpublished = {arXiv:2510.10257},
}

@inproceedings{gao2025easysplat,
  author    = {Gao, Ao and Guo, Luosong and Chen, Tao and Wang, Zhao and Tai, Ying and Yang, Jian and Zhang, Zhenyu},
  title     = {EasySplat: View-Adaptive Learning Makes {3D} Gaussian Splatting Easy},
  year      = {2025},
  booktitle = {ICME},
  publisher = {IEEE},
  address   = {Nantes, France},
  pages     = {1--6},
}

@inproceedings{wang2024group,
  author    = {Wang, Chengbo and Ma, Guozheng and Xue, Yifei and Lao, Yizhen},
  title     = {Faster and Better {3D} Splatting via Group Training},
  year      = {2025},
  booktitle = {ICCV},
  publisher = {IEEE},
  address   = {Honolulu, HI, USA},
  pages     = {27968--27977},
}

@inproceedings{lan2025gs2,
  author    = {Lan, Lei and Shao, Tianjia and Lu, Zixuan and Zhang, Yu and Jiang, Chenfanfu and Yang, Yin},
  title     = {{3DGS}$^2$: Near Second-order Converging {3D} Gaussian Splatting},
  booktitle = {SIGGRAPH},
  year      = {2025},
  note      = {arXiv:2501.13975},
  publisher = {ACM},
  address   = {Vancouver, BC, Canada},
  numpages  = {10},
}

@misc{zhang2026fastconverging,
  author       = {Zhang, Ziyu and Liu, Tianle and Tu, Diantao and Shen, Shuhan},
  title        = {Fast Converging {3D} Gaussian Splatting for 1-Minute Reconstruction},
  year         = {2026},
  howpublished = {arXiv:2601.19489},
}

@inproceedings{feng2024flashgs,
  author    = {Feng, Guofeng and Chen, Siyan and Fu, Rong and Liao, Zimu and Wang, Yi and Liu, Tao and Pei, Zhilin and Li, Hengjie and Zhang, Xingcheng and Dai, Bo},
  title     = {FlashGS: Efficient {3D} Gaussian Splatting for Large-scale and High-resolution Rendering},
  year      = {2025},
  booktitle = {CVPR},
  publisher = {IEEE},
  address   = {Nashville, TN, USA},
  pages     = {26652--26662},
}

@misc{yuan2025hardware,
  author       = {Yuan, Yitian and He, Qianyue},
  title        = {Efficient Differentiable Hardware Rasterization for {3D} Gaussian Splatting},
  year         = {2025},
  howpublished = {arXiv:2505.18764},
}

@inproceedings{kheradmand2025stochasticsplats,
  author    = {Kheradmand, Shakiba and Vicini, Delio and Kopanas, Georgios and Lagun, Dmitry and Yi, Kwang Moo and Matthews, Mark and Tagliasacchi, Andrea},
  title     = {StochasticSplats: Stochastic Rasterization for Sorting-Free {3D} Gaussian Splatting},
  year      = {2025},
  booktitle = {ICCV},
  publisher = {IEEE},
  address   = {Honolulu, HI, USA},
  pages     = {26326--26335},
}

@inproceedings{hou2024sortfree,
  author    = {Hou, Qiqi and Rauwendaal, Randall and Li, Zifeng and Le, Hoang and Farhadzadeh, Farzad and Porikli, Fatih and Bourd, Alexei and Said, Amir},
  title     = {Sort-free Gaussian Splatting via Weighted Sum Rendering},
  booktitle = {ICLR},
  year      = {2025},
  note      = {arXiv:2410.18931},
  publisher = {OpenReview.net},
  address   = {Singapore},
  numpages  = {17},
}

@article{hahlbohm2024hybrid,
  author    = {Hahlbohm, Florian and Friederichs, Fabian and Weyrich, Tim and Franke, Linus and Kappel, Moritz and Castillo, Susana and Stamminger, Marc and Eisemann, Martin and Magnor, Marcus},
  title     = {Efficient Perspective-Correct {3D} Gaussian Splatting Using Hybrid Transparency},
  journal   = {Computer Graphics Forum},
  year      = {2025},
  volume    = {44},
  number    = {2},
  doi       = {10.1111/cgf.70014},
  articleno = {e70014},
  numpages  = {12},
}

@inproceedings{gao2025proxygs,
  author    = {Gao, Yuanyuan and Gong, Yuning and Liu, Yifei and Li, Jingfeng and Zhong, Zhihang and Zhang, Dingwen and Zhang, Yanci and Xu, Dan and Sun, Xiao},
  title     = {Proxy-{GS}: Efficient {3D} Gaussian Splatting via Proxy Mesh},
  year      = {2026},
  booktitle = {CVPR},
  publisher = {IEEE/CVF},
  address   = {Denver, CO, USA},
  pages     = {7330--7339},
}

@misc{lichtfeld,
  author       = {{LichtFeld Studio}},
  title        = {{LichtFeld Studio}},
  year         = {2025},
  howpublished = {\url{https://github.com/MrNeRF/LichtFeld-Studio}},
  note         = {Software},
}

@inproceedings{yang2026arf,
  author    = {Yang, Yixin and Wu, Bojian and Zhou, Yang and Huang, Hui},
  title     = {Augmented Radiance Field: A General Framework for Enhanced {Gaussian} Splatting},
  booktitle = {ICLR},
  year      = {2026},
  note      = {arXiv:2602.19916},
  publisher = {OpenReview.net},
  address   = {Rio de Janeiro, Brazil},
  numpages  = {22},
}

@inproceedings{yang2024specgaussian,
  author    = {Yang, Ziyi and Gao, Xinyu and Sun, Yangtian and Huang, Yihua and Lyu, Xiaoyang and Zhou, Wen and Jiao, Shaohui and Qi, Xiaojuan and Jin, Xiaogang},
  title     = {Spec-{Gaussian}: Anisotropic View-Dependent Appearance for {3D} {Gaussian} Splatting},
  booktitle = {NeurIPS},
  year      = {2024},
  note      = {arXiv:2402.15870},
  publisher = {Curran Associates, Inc.},
  address   = {Vancouver, BC, Canada},
  pages     = {61192--61216},
}

@inproceedings{disario2025sphericalvoronoi,
  author    = {Di Sario, Francesco and Rebain, Daniel and Verbin, Dor and Grangetto, Marco and Tagliasacchi, Andrea},
  title     = {Spherical {Voronoi}: Directional Appearance as a Differentiable Partition of the Sphere},
  year      = {2026},
  booktitle = {CVPR},
  publisher = {IEEE/CVF},
  address   = {Denver, CO, USA},
  pages     = {22529--22538},
}

@inproceedings{liu2026ubs,
  author    = {Liu, Rong and Gao, Zhongpai and Planche, Benjamin and Chen, Meida and Nguyen, Van Nguyen and Zheng, Meng and Choudhuri, Anwesa and Chen, Terrence and Wang, Yue and Feng, Andrew and Wu, Ziyan},
  title     = {Universal Beta Splatting},
  booktitle = {ICLR},
  year      = {2026},
  note      = {arXiv:2510.03312},
  publisher = {OpenReview.net},
  address   = {Rio de Janeiro, Brazil},
  numpages  = {20},
}

@inproceedings{lu2024scaffoldgs,
  author    = {Lu, Tao and Yu, Mulin and Xu, Linning and Xiangli, Yuanbo and Wang, Limin and Lin, Dahua and Dai, Bo},
  title     = {Scaffold-{GS}: Structured {3D} {Gaussians} for View-Adaptive Rendering},
  booktitle = {CVPR},
  year      = {2024},
  note      = {arXiv:2312.00109},
  publisher = {IEEE},
  address   = {Seattle, WA, USA},
  pages     = {20654--20664},
}

@misc{kingma2015adam,
  author       = {Diederik P. Kingma and
                  Jimmy Ba},
  title        = {Adam: {A} Method for Stochastic Optimization},
  year         = {2015},
  url          = {https://doi.org/10.48550/arXiv.1412.6980},
  doi          = {10.48550/ARXIV.1412.6980},
  howpublished = {International Conference on Learning Representations (ICLR), San Diego, CA, USA},
}

@misc{higs,
  title        = {{HiGS}: A Hierarchical Rendering Architecture for Real-Time {3D} {Gaussian} Splatting},
  author       = {Paj{\k{a}}k, Dawid and Bisson, Martin and Lima, Rodolfo},
  year         = {2026},
  howpublished = {arXiv:2606.00352},
}

@inproceedings{liao2026litegs,
  author    = {Kaimin Liao and
                  Hua Wang and
                  Zhi Chen and
                  Luchao Wang and
                  Yaohua Tang},
  title     = {LiteGS: {A} High-Performance Framework to Train 3DGS in Subminutes
                  via System and Algorithm Codesign},
  booktitle = {Computer Vision - {ECCV} 2026 - 19th European Conference, Malm{\"{o}},
                  Sweden, September 8-12, 2026, Proceedings, Part {XLII}},
  pages     = {574--591},
  year      = {2026},
  url       = {https://doi.org/10.1007/978-3-032-37422-6\_33},
  doi       = {10.1007/978-3-032-37422-6\_33},
  bibsource = {dblp computer science bibliography, https://dblp.org},
  publisher = {Springer},
  address   = {Malm{\"{o}}, Sweden},
}

@misc{fusedssim,
  author       = {Goel, Rahul},
  title        = {{fused-ssim}: Fully Fused Differentiable {SSIM}},
  howpublished = {\url{https://github.com/rahul-goel/fused-ssim}},
  year         = {2024},
  note         = {Commit a7c48d6},
}

@misc{luo2026roofgs,
  author       = {Luo, Yang and Gong, Yan and Gao, Yongsheng and Zhao, Jie},
  title        = {{RoofGS}: Roofline-Guided End-to-End Acceleration of {3D} {Gaussian} Splatting},
  year         = {2026},
  howpublished = {arXiv:2608.15785},
}

\end{document}


\maketitle

\appendix
\section{Reproducibility notes}
\label{sec:supp:repro}
\TODO{}

\subsection{Device configurations}
\label{sec:supp:devices}
Full configurations of the devices listed in the main paper's device table.
\TODO{CPU, RAM, OS, driver, and backend (Vulkan / MoltenVK / ROCm) for
Specs 1--4; add the fifth device.}

%

\section{Additional figures and tables}
\label{sec:appendix:extra}

\begin{figure*}[!p]
  \centering
  \begin{subfigure}{0.48\textwidth}
    \centering
    \includegraphics[width=\linewidth]{abl_curve_psnr_3060.pdf}
    \caption{PSNR, RTX 3060}
  \end{subfigure}\hfill
  \begin{subfigure}{0.48\textwidth}
    \centering
    \includegraphics[width=\linewidth]{abl_curve_ssim_3060.pdf}
    \caption{SSIM, RTX 3060}
  \end{subfigure}

  \vspace{6pt}
  \begin{subfigure}{0.48\textwidth}
    \centering
    \includegraphics[width=\linewidth]{abl_curve_psnr_3090.pdf}
    \caption{PSNR, RTX 3090}
  \end{subfigure}\hfill
  \begin{subfigure}{0.48\textwidth}
    \centering
    \includegraphics[width=\linewidth]{abl_curve_ssim_3090.pdf}
    \caption{SSIM, RTX 3090}
  \end{subfigure}
  \caption{Ablation ladder on the RTX~3060 (top) and RTX~3090 (bottom), the
  two devices not shown in the main paper. Protocol as in the main-paper
  ladder figure: scene-balanced mean over all fifteen evaluation scenes,
  full-resolution eval against the full held-out split, components removed one
  at a time from the full system, dots at 30k/60k/80k steps.
  Note the x-axis span differs per device.}
  \label{fig:supp:ladder}
\end{figure*}

\begin{figure*}[!p]
  \centering
  \begin{subfigure}{0.24\textwidth}
    \centering\includegraphics[width=\linewidth]{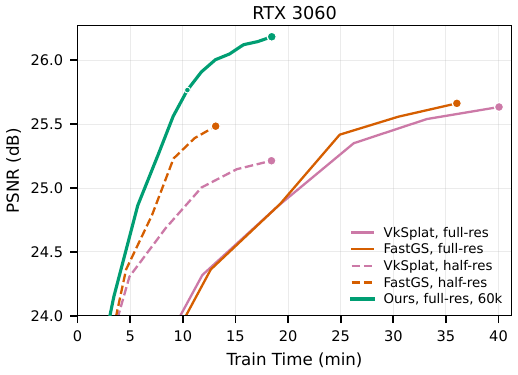}
    \caption{RTX 3060}
  \end{subfigure}\hfill
  \begin{subfigure}{0.24\textwidth}
    \centering\includegraphics[width=\linewidth]{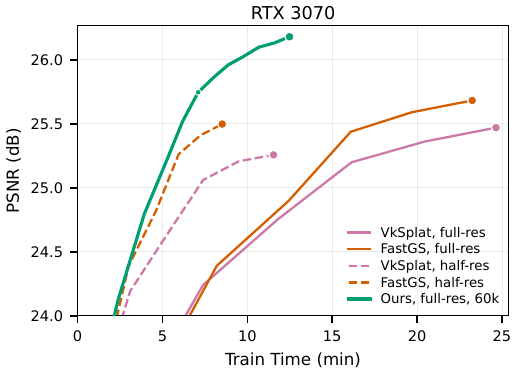}
    \caption{RTX 3070}
  \end{subfigure}\hfill
  \begin{subfigure}{0.24\textwidth}
    \centering\includegraphics[width=\linewidth]{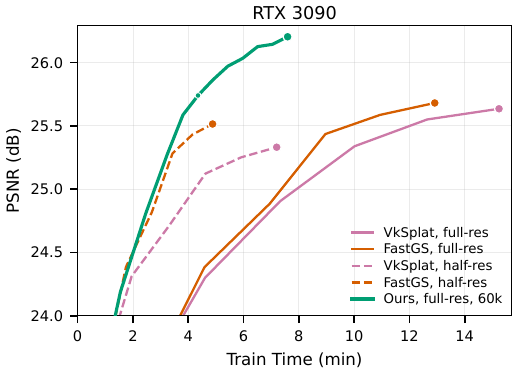}
    \caption{RTX 3090}
  \end{subfigure}\hfill
  \begin{subfigure}{0.24\textwidth}
    \centering\includegraphics[width=\linewidth]{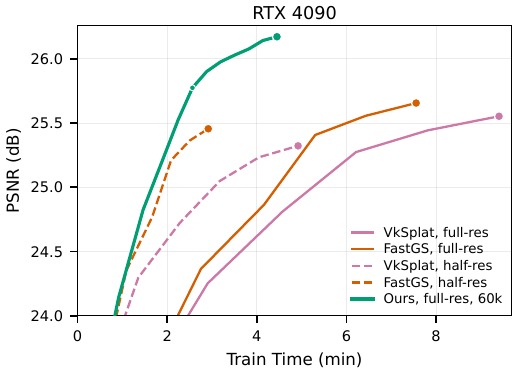}
    \caption{RTX 4090}
  \end{subfigure}
  \caption{Training resolution across systems, on every device
  measured. Dashed curves are the same systems trained at half resolution; every
  model is rendered at full resolution to be scored. The ordering is the same on
  all four devices: half-resolution training is cheaper and settles lower, and
  the gap does not close with more time.}
  \label{fig:sampling_curve}
\end{figure*}

\begin{figure*}[!p]
  \centering
  \begin{subfigure}{0.325\textwidth}
    \centering\includegraphics[width=\linewidth]{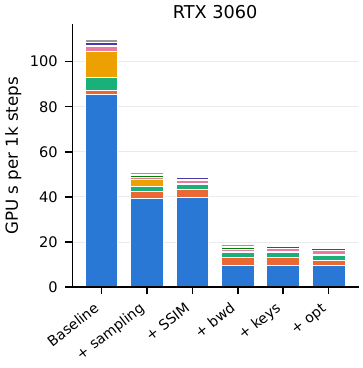}
  \end{subfigure}\hfill
  \begin{subfigure}{0.325\textwidth}
    \centering\includegraphics[width=\linewidth]{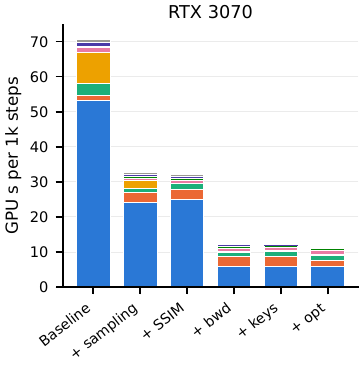}
  \end{subfigure}\hfill
  \begin{subfigure}{0.325\textwidth}
    \centering\includegraphics[width=\linewidth]{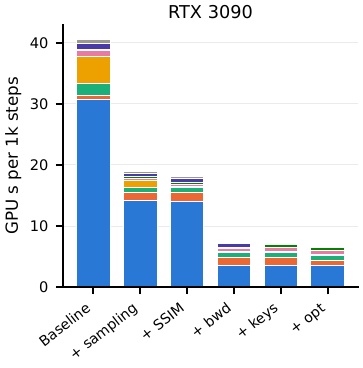}
  \end{subfigure}

  \vspace{2pt}
  \includegraphics[width=\textwidth]{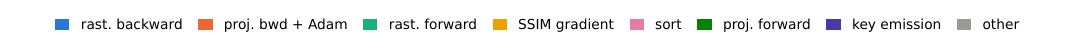}
  \caption{Per-stage GPU time down the ablation ladder, seconds per 1k training
  steps, scene-balanced over all fifteen evaluation scenes. Note the per-device
  y-scale. The two halves of the rung merged in the main paper are separated
  here: \mbox{$+$~Fused Keys} and \mbox{$+$~Opt}.}
  \label{fig:supp:stages}
\end{figure*}

\begin{figure*}[!p]
  \centering
  \includegraphics[width=\linewidth]{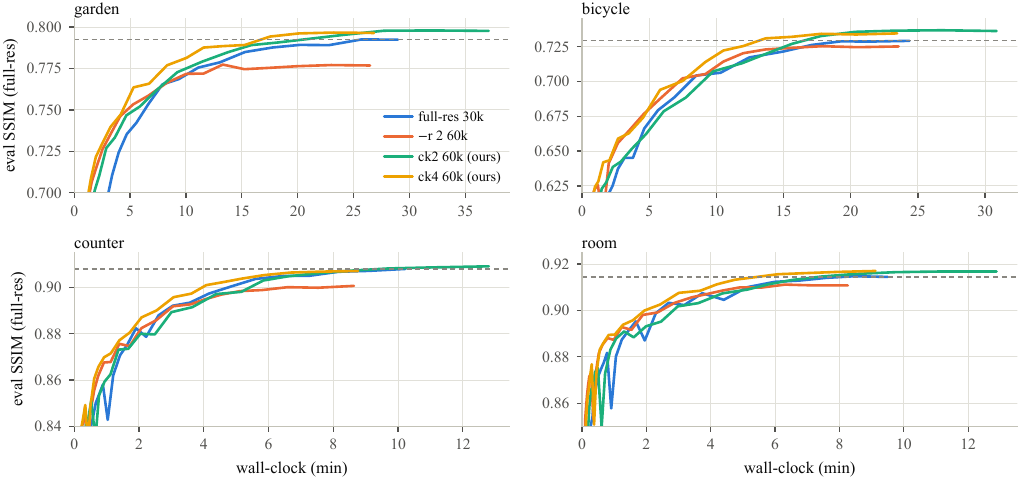}
  \caption{Held-out SSIM (native full-resolution eval, full held-out set at
  30 log-spaced steps) vs.\ wall-clock, per scene, on the RTX~3070.
  \emph{full-res}: full-resolution training; \emph{$-r$~2}: half-resolution
  training; \emph{ck2}/\emph{ck4}: \nicefrac{1}{2}- and \nicefrac{1}{4}-strided
  pixel sampling (\cref{sec:method}); the number after each label is the step
  budget. Dashed line: the full-resolution baseline's final quality on that
  scene.}
  \label{fig:supp:facets}
\end{figure*}

\begin{figure}[!p]
  \centering
  \includegraphics[width=\linewidth]{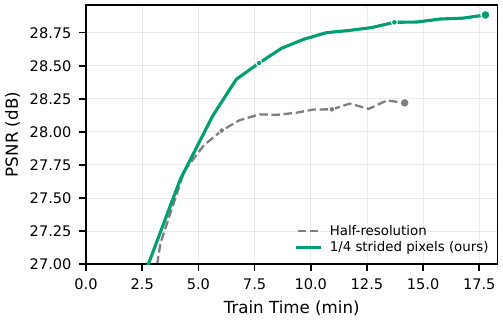}
  \caption{Strided pixel sampling against half-resolution
  training with the rest of the system unchanged, RTX~3070, seven Mip-NeRF~360
  scenes, \num{80000} steps. Both are rendered at full resolution to be scored.}
  \label{fig:ck4_vs_r2}
\end{figure}

%
%

\begin{table*}[t]\centering\scriptsize
\setlength{\tabcolsep}{4pt}
\caption{Per-stage GPU time down the ablation ladder, in seconds per 1k training steps, scene-balanced over all fifteen evaluation scenes. Per-step normalisation is necessary because the arms run different budgets (30k--80k). The two halves of the merged rung are separated here: \emph{+ Fused Keys} empties the key-emission stage but returns the saving through a slower sort and a heavier projection, while \emph{+ Opt} accounts for the whole net gain, in the fused projection-backward and optimizer step. Stages listed here account for roughly 98\% of measured training wall-clock; sub-timers that re-report other stages are excluded.}

\begin{tabular}{@{}ll rrrrrrr r r@{}}
\toprule
GPU & Configuration & rast.\ bwd & proj.\ bwd$+$Adam & rast.\ fwd & SSIM grad & sort & proj.\ fwd & keys & other & total \\
\midrule
\multirow{6}{*}{RTX3060} & Baseline & 85.50 & 1.67 & 5.70 & 11.60 & 2.25 & 0.36 & 1.45 & 1.14 & \textbf{109.67} \\
 & + Pixel Sampling & 39.22 & 3.34 & 2.22 & 2.83 & 1.14 & 0.63 & 0.71 & 0.60 & \textbf{50.68} \\
 & + Tile SSIM & 39.79 & 3.50 & 2.26 & 0.37 & 1.15 & 0.64 & 0.74 & 0.55 & \textbf{49.01} \\
 & + Rast. Bwd & 9.75 & 3.41 & 2.19 & 0.37 & 1.13 & 0.63 & 0.75 & 0.50 & \textbf{18.72} \\
 & + Fused Keys & 9.72 & 3.39 & 2.18 & 0.37 & 1.59 & 0.83 & 0.02 & 0.37 & \textbf{18.48} \\
 & + Opt & 9.75 & 2.20 & 2.20 & 0.37 & 1.62 & 0.83 & 0.02 & 0.34 & \textbf{17.33} \\
\addlinespace[2pt]
\multirow{6}{*}{RTX3070} & Baseline & 53.24 & 1.34 & 3.55 & 8.62 & 1.52 & 0.29 & 1.19 & 0.81 & \textbf{70.55} \\
 & + Pixel Sampling & 24.18 & 2.72 & 1.37 & 2.06 & 0.78 & 0.51 & 0.59 & 0.46 & \textbf{32.66} \\
 & + Tile SSIM & 25.09 & 2.88 & 1.44 & 0.25 & 0.78 & 0.52 & 0.61 & 0.43 & \textbf{32.01} \\
 & + Rast. Bwd & 5.92 & 2.76 & 1.33 & 0.25 & 0.76 & 0.51 & 0.61 & 0.38 & \textbf{12.53} \\
 & + Fused Keys & 5.96 & 2.77 & 1.34 & 0.26 & 1.03 & 0.70 & 0.02 & 0.28 & \textbf{12.36} \\
 & + Opt & 5.95 & 1.78 & 1.34 & 0.26 & 1.05 & 0.69 & 0.02 & 0.25 & \textbf{11.34} \\
\addlinespace[2pt]
\multirow{6}{*}{RTX3090} & Baseline & 30.65 & 0.67 & 2.04 & 4.47 & 0.88 & 0.20 & 1.04 & 0.58 & \textbf{40.52} \\
 & + Pixel Sampling & 14.13 & 1.33 & 0.81 & 1.12 & 0.46 & 0.30 & 0.51 & 0.36 & \textbf{19.01} \\
 & + Tile SSIM & 14.07 & 1.38 & 0.81 & 0.16 & 0.46 & 0.30 & 0.54 & 0.34 & \textbf{18.06} \\
 & + Rast. Bwd & 3.49 & 1.36 & 0.78 & 0.16 & 0.46 & 0.30 & 0.51 & 0.29 & \textbf{7.35} \\
 & + Fused Keys & 3.53 & 1.36 & 0.79 & 0.17 & 0.61 & 0.51 & 0.02 & 0.21 & \textbf{7.20} \\
 & + Opt & 3.53 & 0.88 & 0.80 & 0.17 & 0.62 & 0.52 & 0.02 & 0.19 & \textbf{6.73} \\
\bottomrule
\end{tabular}
\label{tab:stage_breakdown}
\end{table*}

\begin{table*}[t]\centering\small
\setlength{\tabcolsep}{3pt}
\caption{Training resolution across systems. Each competitor is trained at full resolution and again at half resolution with nothing else changed; every model is rendered at full resolution to be scored, so no upsampling enters the measurement. All values pool the fifteen evaluation scenes -- seven Mip-NeRF~360 and eight Tanks \& Temples 4K, each scene weighted equally -- so the level describes the suite rather than either dataset. Competitors run \num{30000} steps and ours is quoted at the same budget. The right-hand columns are training wall-clock on the card named. The RTX~4090 panel is in the supplementary material. Shading marks the best three in each column.}

\begin{tabular}{@{}l rrr rrr@{}}
\toprule
Configuration & PSNR$\uparrow$ & SSIM$\uparrow$ & LPIPS$\downarrow$ & RTX3060 & RTX3070 & RTX3090 \\
\midrule
VkSplat, $-r$\,1 & \cellcolor{rankthree}25.58 & \cellcolor{ranktwo}0.825 & 0.294 & 40m \phantom{0}3s & 24m 39s & 15m 14s \\
FastGS, $-r$\,1 & \cellcolor{ranktwo}25.67 & 0.820 & 0.304 & 36m \phantom{0}3s & 23m 15s & 12m 55s \\
VkSplat, $-r$\,2 & 25.29 & \cellcolor{rankthree}0.821 & \cellcolor{ranktwo}0.269 & \cellcolor{rankthree}18m 25s & \cellcolor{rankthree}11m 32s & \cellcolor{rankthree}\phantom{0}7m 12s \\
FastGS, $-r$\,2 & 25.49 & 0.817 & \cellcolor{rankthree}0.291 & \cellcolor{ranktwo}13m \phantom{0}7s & \cellcolor{ranktwo}\phantom{0}8m 31s & \cellcolor{ranktwo}\phantom{0}4m 53s \\
\midrule
\textbf{Ours, $-r$\,1 (30k)} & \cellcolor{rankone}\textbf{25.76} & \cellcolor{rankone}\textbf{0.826} & \cellcolor{rankone}\textbf{0.268} & \cellcolor{rankone}\textbf{10m 26s} & \cellcolor{rankone}\textbf{\phantom{0}7m \phantom{0}6s} & \cellcolor{rankone}\textbf{\phantom{0}4m 21s} \\
\bottomrule
\end{tabular}
\label{tab:sampling}
\end{table*}

\begin{table}[t]\centering\scriptsize
\setlength{\tabcolsep}{2.2pt}
\caption{Per-scene results on a MacBook Air with the Apple M4 chip (10-core CPU, 8-core GPU). Training is at half resolution; both metrics are scored at full resolution against the full held-out split, as everywhere else in the paper, so these numbers sit on the same scale as the GPU results and only the training resolution differs. 60k steps; training wall-clock only.}

\begin{tabular*}{\columnwidth}{@{\extracolsep{\fill}}l rr r@{}}
\toprule
Scene & PSNR$\uparrow$ & SSIM$\uparrow$ & Apple M4 \\
\midrule
\multicolumn{4}{@{}l}{\textit{Mip-NeRF 360}} \\
bicycle & 22.88 & 0.681 & 59m 52s \\
bonsai & 31.93 & 0.931 & 25m 51s \\
counter & 28.92 & 0.902 & 29m \phantom{0}8s \\
garden & 25.94 & 0.780 & 90m 54s \\
kitchen & 30.80 & 0.912 & 49m 58s \\
room & 31.83 & 0.913 & 28m 21s \\
stump & 26.46 & 0.801 & 66m 17s \\
\addlinespace[1pt]
\textbf{Mean} & \textbf{28.39} & \textbf{0.845} & \textbf{50m \phantom{0}3s} \\
\midrule
\multicolumn{4}{@{}l}{\textit{Tanks \& Temples 4K}} \\
caterpillar & 22.14 & 0.708 & 31m 35s \\
family & 26.48 & 0.855 & 39m 15s \\
francis & 28.08 & 0.871 & 27m \phantom{0}2s \\
horse & 25.01 & 0.864 & 31m 38s \\
ignatius & 21.52 & 0.687 & 54m 58s \\
lighthouse & 21.84 & 0.816 & 28m 46s \\
train & 21.49 & 0.764 & 27m 28s \\
truck & 24.51 & 0.823 & 42m 43s \\
\addlinespace[1pt]
\textbf{Mean} & \textbf{23.88} & \textbf{0.799} & \textbf{35m 26s} \\
\bottomrule
\end{tabular*}
\label{tab:supp:macbook}
\end{table}

\section{Additional measurements and studies}
\label{sec:supp:extra}

\subsection{Depth dependence of the backward kernels}
\TODO{}

\subsection{Per-scene end-to-end curves}
\label{sec:supp:curves}

\Cref{fig:supp:facets} breaks the mean-over-scenes curves of the main paper
down per scene.

\subsection{Pixel Sampling on every GPU}
\label{sec:supp:sampling}

\cref{tab:sampling} reports the training-resolution comparison.
\cref{fig:sampling_curve} shows it as quality against wall-clock, one panel per
device, so the shape of each configuration's approach can be read as well as
where it ends. The arms, the fifteen-scene
panel, the eval protocol and the axis crop are identical; each panel sets its
own y-range because the devices differ in how far they get.

\subsection{Ablation ladder on the other two GPUs}
\label{sec:supp:ladder}

The main paper shows the ablation ladder on an RTX~3070.
\cref{fig:supp:ladder} repeats it on the RTX~3060 and the RTX~3090.
The arms, the fifteen-scene panel, the eval protocol and the axis crops are
identical, and the two metrics share a y-range with the main-paper panels, so
the three devices can be read against one another directly.

\subsection{Where the time goes, stage by stage}
\label{sec:supp:stages}

Every run in the ablation campaign carries the Vulkan stage timers, so the
ladder can be decomposed into the stage each rung actually moves
(\cref{fig:supp:stages}, exact figures in \cref{tab:stage_breakdown}).
Times are normalised to GPU seconds per 1k training steps: the arms run
budgets from 30k to 80k steps, so raw totals would compare unequal amounts of
work. The stages shown account for roughly \SI{98}{\percent} of measured
training wall-clock.

Two readings do not survive the end-to-end numbers. First, the
\emph{+~Tile SSIM} rung looks slightly slower in total, but its own stage falls
by \SI{88}{\percent} (\SI{2.06}{\second} to \SI{0.25}{\second} per 1k steps on
the RTX~3070); the increase is in rasterize-backward, and it follows that arm
carrying more Gaussians, not the loss. Second, fused key emission does empty
the key-emission stage --- \SI{0.61}{\second} to \SI{0.02}{\second} per 1k steps
--- but the sort implementation it ships with returns the saving through a
slower sort and a heavier projection forward, leaving the rung net-neutral
(\SI{12.53}{\second} to \SI{12.36}{\second}). The net gain of the last rung is
culling, in the fused projection-backward and optimizer step, and it is
consistent across all three cards.

\subsection{Apple silicon, per scene}
\label{sec:supp:macbook}

\Cref{tab:supp:macbook} breaks the MacBook Air M4 summary in the main paper
down per scene. The spread within a dataset is wide --- on Mip-NeRF~360 the
indoor scenes finish in under half an hour while garden takes an hour and a
half --- so the dataset means should be read with the per-scene times in view.

\subsection{Multi-view gradient batching (exploratory)}
\TODO{}

\subsection{Negative results, with mechanisms}
\TODO{}

\bibliographystyle{ACM-Reference-Format}
\bibliography{main}